\documentclass[twocolumn]{aastex61}
\pdfoutput=1 
\usepackage{amsmath,amstext}
\usepackage[T1]{fontenc}
\usepackage{apjfonts} 
\usepackage[figure,figure*]{hypcap}

\usepackage{longtable}
\usepackage{cancel}	
\usepackage{color}
\usepackage{graphicx, subfigure}
\usepackage{placeins}
\usepackage{amsmath}
\usepackage{tabulary}

\def\lesssim{\mathrel{\hbox{\rlap{\hbox{\lower4pt\hbox{$\sim$}}}\hbox{$<$}}}}
\def\gtrsim{\mathrel{\hbox{\rlap{\hbox{\lower4pt\hbox{$\sim$}}}\hbox{$>$}}}}

\newcommand{\ips}{\ensuremath{i_{\rm P1}}}

\newcommand{\msun}{\mbox{M$_{\odot}$}}

\newcommand{\kms}{\mbox{$\rm{\,km\,s^{-1}}$}}

\def\ergcm2s{erg\,cm$^{-2}$\,s$^{-1}$}

\shorttitle{ATLAS17aeu}
\shortauthors{Stalder et al.}

\begin{document}

\title{Observations of the GRB afterglow ATLAS17aeu and its possible association with GW170104}

\author{B. Stalder}
\affiliation{Institute for Astronomy, University of Hawaii, 2680 Woodlawn Drive, Honolulu, HI 96822}
\email{stalder@ifa.hawaii.edu}
\author{J. Tonry}
\affiliation{Institute for Astronomy, University of Hawaii, 2680 Woodlawn Drive, Honolulu, HI 96822}
\author{S. J. Smartt}
\affiliation{Astrophysics Research Centre, School of Mathematics and Physics, Queens University Belfast, Belfast BT7 1NN, UK}
\author{M. Coughlin}
\affiliation{Department of Physics, Harvard University, Cambridge, MA 02138, USA}
\author{K. C. Chambers}
\affiliation{Institute for Astronomy, University of Hawaii, 2680 Woodlawn Drive, Honolulu, HI 96822}
\author{C. W. Stubbs}
\affiliation{Department of Physics, Harvard University, Cambridge, MA 02138, USA}
\author{T.-W. Chen}
\affiliation{Max-Planck-Institut f{\"u}r Extraterrestrische Physik, Giessenbachstra\ss e 1, 85748, Garching, Germany}
\author{E. Kankare}
\affiliation{Astrophysics Research Centre, School of Mathematics and Physics, Queens University Belfast, Belfast BT7 1NN, UK}
\author{K. W. Smith}
\affiliation{Astrophysics Research Centre, School of Mathematics and Physics, Queens University Belfast, Belfast BT7 1NN, UK}
\author{L. Denneau}
\affiliation{Institute for Astronomy, University of Hawaii, 2680 Woodlawn Drive, Honolulu, HI 96822}
\author{A. Sherstyuk}
\affiliation{Institute for Astronomy, University of Hawaii, 2680 Woodlawn Drive, Honolulu, HI 96822}
\author{A. Heinze}
\affiliation{Institute for Astronomy, University of Hawaii, 2680 Woodlawn Drive, Honolulu, HI 96822}
\author{H. Weiland}
\affiliation{Institute for Astronomy, University of Hawaii, 2680 Woodlawn Drive, Honolulu, HI 96822}
\author{A. Rest}
\affiliation{Space Telescope Science Institute, 3700 San Martin Drive, Baltimore, MD 21218, USA}
\author{D. R. Young}
\affiliation{Astrophysics Research Centre, School of Mathematics and Physics, Queens University Belfast, Belfast BT7 1NN, UK}
\author{M. E. Huber}
\affiliation{Institute for Astronomy, University of Hawaii, 2680 Woodlawn Drive, Honolulu, HI 96822}
\author{H. Flewelling}
\affiliation{Institute for Astronomy, University of Hawaii, 2680 Woodlawn Drive, Honolulu, HI 96822}
\author{T. Lowe}
\affiliation{Institute for Astronomy, University of Hawaii, 2680 Woodlawn Drive, Honolulu, HI 96822}
\author{E. A. Magnier}
\affiliation{Institute for Astronomy, University of Hawaii, 2680 Woodlawn Drive, Honolulu, HI 96822}
\author{A. S. B. Schultz}
\affiliation{Institute for Astronomy, University of Hawaii, 2680 Woodlawn Drive, Honolulu, HI 96822}
\author{C. Waters}
\affiliation{Institute for Astronomy, University of Hawaii, 2680 Woodlawn Drive, Honolulu, HI 96822}
\author{R. Wainscoat}
\affiliation{Institute for Astronomy, University of Hawaii, 2680 Woodlawn Drive, Honolulu, HI 96822}
\author{M. Willman}
\affiliation{Institute for Astronomy, University of Hawaii, 2680 Woodlawn Drive, Honolulu, HI 96822}
\author{D. E. Wright}
\affiliation{School of Physics and Astronomy, University of Minnesota,  116 Church Street SE, Minneapolis, MN 55455-0149}
\author{J. K. Chu}
\affiliation{Institute for Astronomy, University of Hawaii, 2680 Woodlawn Drive, Honolulu, HI 96822}
\author{D. Sanders }
\affiliation{Institute for Astronomy, University of Hawaii, 2680 Woodlawn Drive, Honolulu, HI 96822}
\author{C. Inserra}
\affiliation{Astrophysics Research Centre, School of Mathematics and Physics, Queens University Belfast, Belfast BT7 1NN, UK}
\author{K. Maguire}
\affiliation{Astrophysics Research Centre, School of Mathematics and Physics, Queens University Belfast, Belfast BT7 1NN, UK}
\author{R. Kotak}
\affiliation{Astrophysics Research Centre, School of Mathematics and Physics, Queens University Belfast, Belfast BT7 1NN, UK}

\begin{abstract}

We report the discovery and multi-wavelength data analysis of 
the peculiar optical transient, ATLAS17aeu. 
This transient was
identified in the skymap of the 
LIGO gravitational wave event GW170104 by our 
ATLAS and Pan-STARRS coverage. 
ATLAS17aeu was discovered 23.1hrs after GW170104
and rapidly faded over the next 3 nights, 
with a spectrum revealing a blue featureless continuum.  
The transient was also detected 
as a fading x-ray source by Swift  
and in the radio at 6 and 15\,GHz. 
A gamma ray burst GRB170105A was detected by 3 
satellites 19.04\,hrs after GW170104 and 4.10\,hrs
before our first optical detection.  
We analyse the multi-wavelength fluxes in the context 
of the known GRB population and 
discuss the observed sky rates of GRBs and their afterglows. 
We find it statistically likely that ATLAS17aeu is an afterglow
associated with GRB170105A, with a chance coincidence ruled out at the 99\% confidence or 2.6$\sigma$. A long, soft GRB within a redshift range of 
$1 \lesssim z  \lesssim 2.9$ would be consistent with 
all the observed  multi-wavelength data. 
The Poisson probability of a chance occurrence of  GW170104 and ATLAS17aeu  is $p=0.04$.  This is the probability of a chance coincidence in 2D sky location and in time. 
These observations indicate that ATLAS17aeu is plausibly a normal GRB afterglow at
significantly higher redshift than the distance constraint
for GW170104 and therefore a chance coincidence. However if a redshift of the 
faint host were to place it  within the GW170104 distance range, then physical 
association with GW170104 should be considered.  
\end{abstract}


\section{Introduction}

The Advanced LIGO experiment began detecting gravitational waves from the merging of black hole (BH) binary systems in 
2015 \citep{theprizepaper}
.  In addition to being the first direct detection of gravitational waves, these first three detections 
showed that remarkably high stellar mass BHs exist, with the highest mass system (GW150914) resulting from the coalescence 
of a  36\,\msun\ and 29\,\msun\ system.  As of yet, no probable electromagnetic counterparts to these events have been found.  However the {\it Fermi} Gamma-ray Burst Monitor (GBM) detected a weak, hard x-ray transient temporally coincident with GW150914 \citep{2016ApJ...826L...6C},  but this has not been confirmed to be physically associated and its astrophysical nature has been disputed \citep{2016ApJ...827L..38G}.

The second science run of the Advanced LIGO experiment (designated O2) started MJD=57722 (2016-11-30) with a short holiday break between MJD=57744 
and MJD=57756.
Shortly after the break, the internal system distributed the alert of a candidate gravitational wave (GW) transient, designated as event G268556 at 2017-01-04 10:11:58.599 UTC or MJD = 57757.42498378. It was later given the name GW170104 after the offline analysis provided very strong confirmation of its astrophysical origin \citep[as presented in][]{gw170104}. 
The 90\% probability area of the associated GW skymap is 2000 square degrees or 5\% of the
sky. Throughout this paper we use 
the LALInference sky map released on 2017 May 02 by LIGO 
i.e.  \texttt{LALInference\_f.fits} 
\citep{2015PhRvD..91d2003V}.  The estimated binary black hole masses are 
$31.2_{-6.0}^{+8.4}$\,\msun\
and 
 $19.4_{-5.9}^{+5.3}$\,\msun\
at a luminosity distance of 
 $880_{-390}^{+450}$\,Mpc corresponding to a redshift of 
 $0.18_{-0.07}^{+0.08}$ \citep{gw170104}. 

As with previous LIGO events, many teams with electromagnetic follow-up facilities observed the corresponding sky map of this event 
\citep[e.g. for a summary of the GW150914 effort see][]{2016ApJ...826L..13A}. 
Here we introduce the Asteroid Terrestrial-impact Last Alert System \citep[ATLAS;][]{2011PASP..123...58T}, a full-time near Earth asteroid survey,  
and describe its early observations of GW170104 in combination with 
our established Pan-STARRS follow-up program  \citep{2016arXiv161205560C,2016MNRAS.462.4094S,2016ApJ...827L..40S}
During the course of our coordinated ATLAS and Pan-STARRS observations, we identified a bright optical transient, ATLAS17aeu, discovered 23.1hrs after GW170104 and observed to be 
rapidly fading over 2\,hrs.  We discuss the nature of this optical transient and its relation to both the gravitational wave source
GW170104 and the gamma ray burst GRB170105A detected by several gamma-ray satellite missions. 
Throughout this paper, we use the same cosmological parameters adopted by LIGO of  
$H_0 = 69\,\kms,\Omega_{\rm M} = 0.31,\Omega_\Lambda =  0.69$. 

\section{Discovery of ATLAS17aeu and follow-up optical observations}
\label{sec:obs}

\begin{figure}
\includegraphics[width=\columnwidth,angle=0]{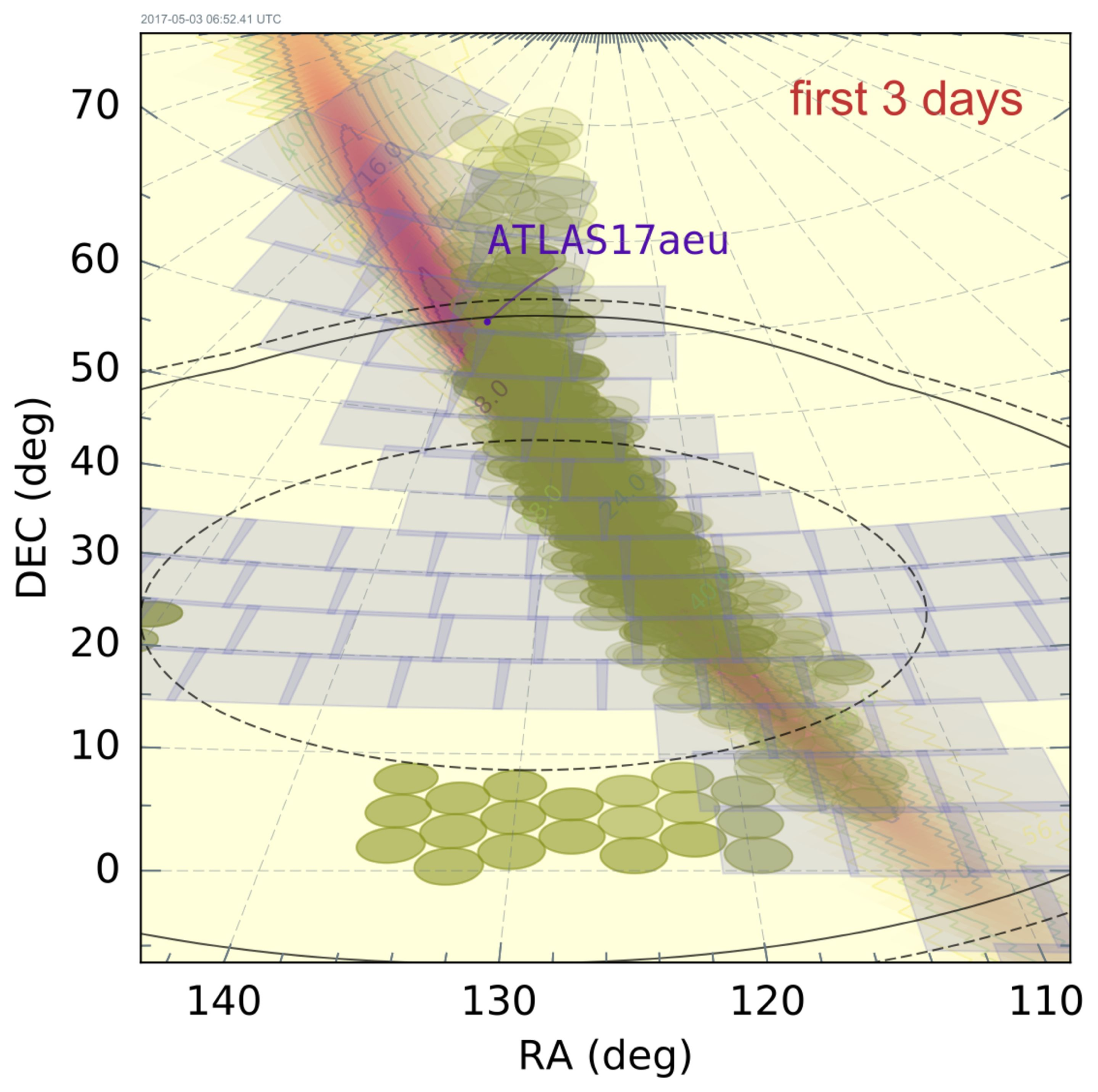}
\caption{
{LALInference likelihood map (\texttt{LALInference\_f.fits})
showing ATLAS (purple squares) and Pan-STARRS1 pointings (green circles). The black solid circle is the best estimate localization of GRB170105A with the black dashed circles representing the Konus-Integral triangulation annulus of
\cite{GCN20406}}
}
\label{fig:skymap}
\end{figure}

The ATLAS system\footnote{fallingstar.com} currently consists of two 0.5-meter f/2 wide-field telescopes \citep{2011PASP..123...58T}, 
of which only the Haleakala unit was in operation at the time of these observations. 
The ATLAS sensor is a single thermoelectrically-cooled STA1600 detector with 1.86 arcsecond per pixel platescale 
(10.56k$\times$10.56k pixels)
giving a 29.2 square degree field of view. 
As stated above, GW170104 was detected on MJD = 57757.42498378 
\citep{gw170104}
and the notification was sent to partners at 57757.70134 (6.6\,hrs later).  Prior to the alert being released, and 
coincident with the GW detection time, 
 ATLAS was observing a declination strip between 
$14^{\circ} < \delta < 35^{\circ}$, cutting through the skymap (Fig.\ref{fig:skymap}) at this time, with the  
first pointing post-detection that encroached on the non-zero probability region taken at MJD=57757.44622 (0.5\,hrs after the GW detection).  One transient was found in the strip during this Hawaiian night of serendipitous survey operations (ATLAS17ace), but it is a known CV candidate with observed activity in CRTS and Pan-STARRS1 going back to 2014 (the object is CSS140914-075541+264619, PS1-14amh).  When the LIGO GW event alert was processed and night subsequently fell in Hawaii, the ATLAS telescope began targeting the fields identified by the GW170104 skymap (Figure \ref{fig:skymap}), using the same cadence as normally employed for asteroid discovery and identification.  
These observations began at  
57758.29696 
or  20.9\,hrs after the signal detection.
The tiling started at the bottom of the banana shaped skymap at 
RA=108.79, DEC=$-7.662$ and continued north-eastward toward the top of the banana, finishing at 57758.46645 
(25.0\,hrs after the signal detection)
at sky position RA=170.55, DEC=72.314. 
Had there been no 6.6\,hr notification delay, 
ATLAS could have been observing within minutes of the LIGO detection as the event was well placed for Hawaiian night time observing. 
During this first night of dedicated observing, ATLAS covered about 42.6\% of the enclosed probability corresponding to a sky-area of 1231 square degrees.
The ATLAS system uses Canon 8--15mm fisheye cameras that cover most of
the visible sky from Haleakala and Mauna Loa.  By co-adding images
immediately following the instant of GW170104 and using an adjacent
$m\sim5.3$ star as a photometric reference, we can place a 3-sigma
limit for any optical transient at the location of ATLAS17aeu of $m>7.5$
for a 1 minute average brightness, $m>10$ for a 1 hour average, and
$m>11$ between the instant of GW170104 and the end of night in Hawaii.

We observed in
a wide-band filter, designated "cyan" or "$c$", which roughly covers the SDSS/Pan-STARRS $g$ and $r$ filters, and maintained our cadence for identifying moving asteroids, which was to observe each footprint 5 or more times (30\,s exposures, slightly dithered) within about an hour of the first observation of each field. The automatic data processing pipeline results in dark-subtracted, sky-flattened images as well as difference images using a static-sky template generated from previous stacked data.  Source extraction is accomplished on the normal and differenced images using DOPHOT \citep{1993PASP..105.1342S} and TPHOT  (a custom written package
for PSF fitting photometry on ATLAS images).  Sources on the difference images are cataloged in a MySQL database and
merged into objects if there are at least 3 detections from the 5 (or more) images. The objects are subject to a set of quality filters, 
a machine learning algorithm and human scanning, similar to those described for Pan-STARRS transient searches 
\citep{2015MNRAS.449..451W,2016MNRAS.462.4094S}.  ATLAS17aeu was discovered in the first image  pass of this region on 57758.41297 (23.7\,hrs after the GW detection) and detected on 7 subsequent overlapping images in total spanning 1.18\,hrs.  It was discovered at  RA=138.30789, DEC=+61.09267 
(09:13:13.89, +61:05:33.6), with a RMS  positional scatter over the 8 images of 0\farcs52. 
Only two other extragalactic transients were identified  on this night.
One is a nuclear transient (ATLAS17ber = PS17em = AT2017aur) coincident with 
the core of a $r=19.1$ galaxy (SDSS J084004.30+584703.1) which showed variable 
activity for the next 70 days and is likely central AGN activity. The other (ATLAS17afb)
is variability of the known QSO (SDSS J092136.23+621552.1) at $z=1.44746$. Neither
of these are likely related to GW170104. 

In parallel we observed the central high probability region with the Pan-STARRS1 (PS1) system \citep{2016arXiv161205560C}, 
similar to our previous GW events. Our joint ATLAS + PS1 strategy is to cover a wide area (several 1000
square degrees) fast with ATLAS (to roughly 19 mag) and the higher probably region (several 100 square degrees) deeper with PS1 (to roughly 21.5 mag). 
We observed a total of 671 square degrees, covering 
43.4\% of the probability sky-area to a more sensitive depth ($\ips \simeq 21.5-22.0$) on the higher probability region
(Figure \ref{fig:skymap}). 
Data were processed and 115 suspected extragalactic transient objects were detected in the first three days after the event as described in \cite{2016MNRAS.462.4094S}.
Of these, we have  spectroscopically confirmed 12 as supernovae or AGNs (5 SNIa, 2 SNIb, 3 SNII, 2 AGNs), with two uncertain classifications dominated by the host galaxy
and a summary paper on PS1 findings in LIGO O2 is
in preparation. 
PS1 also detected ATLAS17aeu on MJD=57758 and 57759
in the \ips\ filter, with the first 
detection just before the ATLAS17aeu point.  We recovered the transient in the ATLAS data first, 
due to our quicker processing speed for the shallower and smaller data rate (a factor of about 10) of ATLAS
compared to PS1.  All photometry values are reported in Table\,\ref{tab:data}. 

We subsequently observed ATLAS17aeu with Gemini North and GMOS \citep{2004PASP..116..425H}, acquiring both $r-$band imaging and optical spectroscopy on MJD = 57761.595 (2017-01-08.595 UT, 4.2\,days after the GW detection). 
These data were taken with the e2V deep depletion (e2vDD) detector array before it was 
decommissioned in February 2017 (within program GN-2016B-Q-2 PI: Chambers). 
The spectrum was taken with the R400 grating covering 4490-8878\,\AA, with a 1\farcs0 slit in seeing conditions of 1\farcs06. Six separate 900s exposures were taken giving a total of 5400s on source. These 2D data were reduced with the IRAF/Gemini package in a 
standard way, with the arc lamp wavelength calibration checked  using sky emission lines.  
 Two methods of spectrum stacking and sky subtraction were used. The first method simply extracted a spectrum from each
single 900s exposure, with sky subtraction in apertures close to the object. The 
second involved 2D image subtraction, alignment and stacking (since the object
was shifted spatially during the exposures). Similar results were achieved in each case.
The imaging sequence was a set of 6$\times$180\,s exposures which were shifted, aligned 
and stacked together. A series of 14 SDSS reference stars within 2 arcmin of the position of ATLAS17aeu were used to photometrically calibrate the stacked image. ATLAS17aeu was clearly detected in the
GMOS images at 
$r=22.77\pm0.17$ (Table\,\ref{tab:data}) and a good signal-to-noise spectrum 
revealed a blue featureless continuum ($S/N\simeq  10$ per pixel).  We applied flux calibration using a standard star with 
the same instrument setup and a telluric absorption 
line correction for O$_2$ using an atmosphere model 
convolved and rebinned to the spectrometer resolution e.g. as described in \cite{2015A&A...579A..40S}.

When the object had faded we again used Gemini North and GMOS to take a deep $r-$band
image to search for a host galaxy and constrain the redshift of the transient (within program GN-2017A-Q-23 PI: Chambers).
On 2017 April 01, we took a series of  $13\times90$\,s images with a midpoint time of MJD=57844.33 (86.9\,days after the GW detection) during the 1170\,s total exposure time. 
By this time, the new Hamamatsu detector array had been installed in the instrument.
We then  centered the GMOS slit on the position of ATLAS17aeu and set the slit angle to 
also cover the closest visible galaxy, which offset from the source by $1\farcs8$.
On the night of MJD=57845.252 (2017-04-02, 87.8\,days after the GW detection), we took a series of  $9\times 968$\,s exposures  
with the R400 grating, again covering 4490-8878\,\AA. 
The observations used the Nod \& Shuffle mode.  
Nod \& Shuffle mode interleaves short observations 
(typically 60 seconds each) of the object and the sky without reading out the CCD.  The charge is shuffled  on the chip 
into storage regions,  resulting in two separate spectra images in a single CCD exposure,
one containing the object and the other
containing only the sky.   
The two spectra images are then subtracted 
from each other which helps 
remove strong sky emission residuals, 
leaving increased Poisson noise at the position of the skylines.  
As the IRAF/Gemini package does not yet function with the Hamamatsu detector array, 
we employed manual reductions for both the imaging and spectroscopic modes.  After subtracting 
the spectra images from each other we then shifted and 
stacked the 9 exposures together.  
The results are described in Section\,\ref{sec:host}

In addition we used Keck II and the DEIMOS spectrometer to take 
spectra with the same slit position as set for the GMOS spectra on the 
night of MJD=57870 (2017-04-27, 113\,days after the GW detection). We took $3\times1200$\,s with the 
R600ZD grating, and a 1\farcs2 slit,  giving a spectral range of 
5550-9839\,\AA\ (with the OG550 blocking filter) 
and a resolution of 4.5\,\AA\ at the center of the range.  
DEIMOS is an 8-detector array and ATLAS17aeu fell on chips 3 and 7 (Video Input 6 and 14). These two detectors 
data were debiased using overscan, flat-fielded and 
extracted in standard fashion.
After correcting for the 2D distortion on these chips, we forced an extraction
of the weak signal close to the position of ATLAS17aeu. 
The results of the imaging and spectroscopy are in Section\,\ref{sec:host}. 

ATLAS17aeu immediately stood out as an unusual and rare transient in the ATLAS and Pan-STARRS1 data streams that occurred within 1 day of GW170104 \citep{GCN20382}. 
The decay rate of 0.71\,mag\,hr$^{-1}$ is faster than typical cataclysmic variable (CV) decline \citep{1975JBAA...86...30B} 
and slower than M-dwarf flares
\citep{2013ApJ...779...18B}. The PS1 images confirmed the fading on the second night of observations when it 
had faded below the ATLAS detection limit. 
During the first year of general ATLAS surveying we typically took 5 dithered exposures per night, separated across approximately 1hr.  PS1 has been observing in a set of quad exposures each night,  separated by about 15mins each, since 2015.  In both these surveys,  we
find many supernovae, M-dwarf and CV like candidates 
\citep{2016ATel.8680....1T}, but fast transients which decline with this rate are rare (see Section\,\ref{sec:disc} for further discussion of rates). Therefore we immediately 
released this information  \citep[in][]{GCN20382}  to the broad LIGO-Virgo EM-followup effort
\cite[see][]{2016ApJ...826L..13A}. This instigated a series of multi-wavelength observations, 
which we summarize in the next section. 

\section{Multi-wavelength observations and GRB170105A}
\label{sec:multiw}

\begin{figure*}
\includegraphics[width=2.0\columnwidth,angle=0]{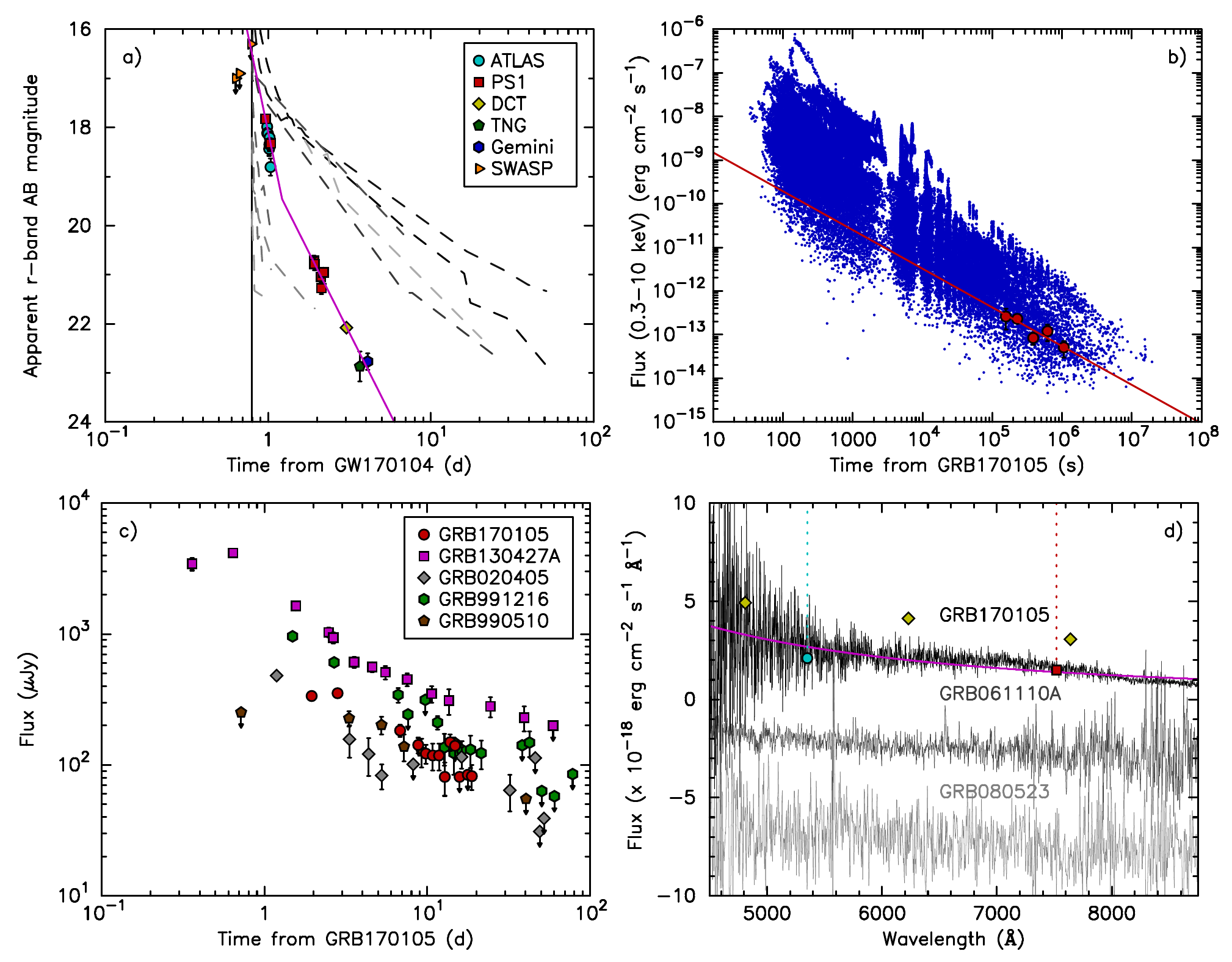}
\caption{
{\bf (a):} The $r-$band lightcurve of ATLAS17aeu with our own data , supplemented with the
photometry reported in LIGO-VIRGO GCNs as listed 
and referenced in Table\,\ref{tab:data}. 
The time of GRB170105A is the vertical black line. All the detections have been colour corrected to observer frame $r-$-band using 
the spectrum in panel (d). 
{\bf (b):} The  x-ray afterglow lightcurves of Swift GRBs with known redshifts from 2005 to present. The ATLAS17aeu fluxes from Swift XRT are
in red. 
{\bf (c):}  The radio fluxes of ATLAS17aeu and other GRBs with radio measurements in the 8-15\,GHz bands. 
{\bf (d):}  The GMOS spectrum of ATLAS17aeu at +3.3\,days after GRB170105A. The 
SED from the $gri$ points of \cite{GCN20397}  at +2.3\,days
are shown for reference. We also show the relative SED
of our photometry (ATLAS cyan and Pan-STARRS \ips\, scaled with the same factor)
at only +4.8\,hr after GRB170105A. This indicates the color
of the afterglow was relatively constant over the first 3
days.  
}
\label{fig:lc}
\end{figure*}

\begin{figure}
\includegraphics[width=\columnwidth,angle=0]{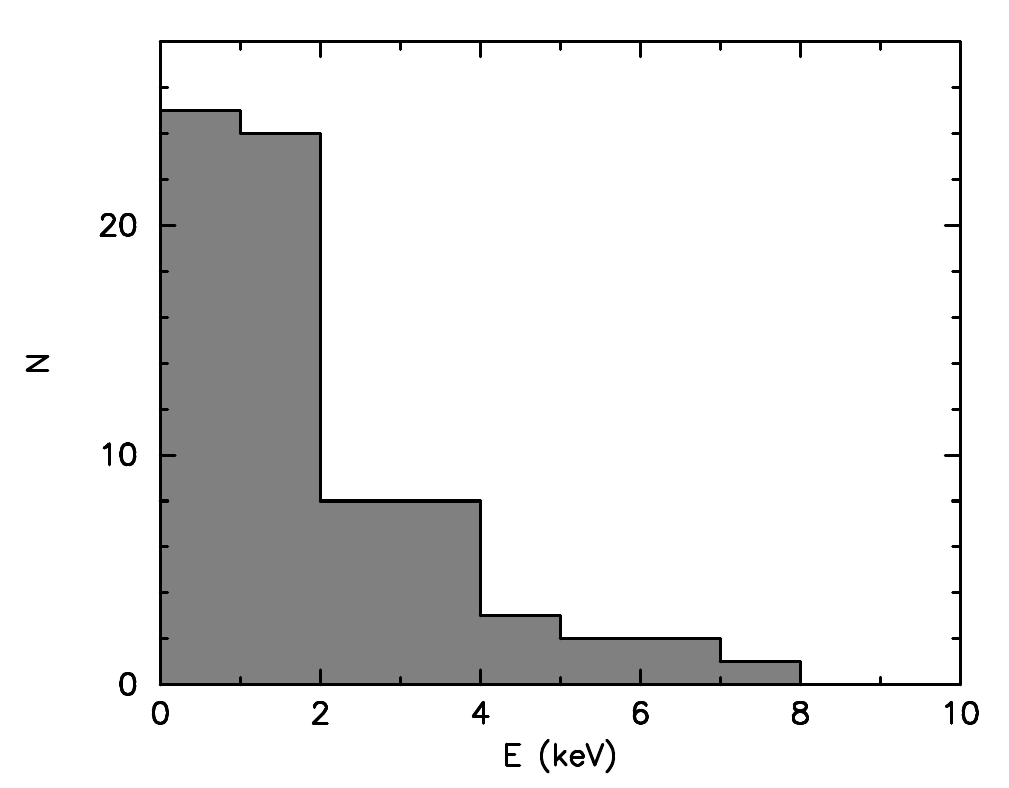}
\caption{A histogram of the energy distribution of individual photons from the 5 epochs with XRT X-ray detection.}
\label{fig:xrt}
\end{figure}

ATLAS17aeu was observed in both the x-ray by Swift \citep{GCN20390}
and the radio by the Arcminute MicroKelvin Imager \cite[AMI;][]{GCN20425} and the Very Large Array 
\citep[VLA][]{GCN20396}. 
\cite{GCN20390} reported an x-ray 
source with Swift (0.3-10\,keV; RA, Dec = 138.3059,+61.0919) on MJD=57760.034 (2.6\,days after the GW detection)
at 4\farcs1$\pm$6\farcs3 from the position of ATLAS17aeu. We downloaded the Swift X-ray Telescope (XRT) 
images from the archive and measured fluxes using the HEASARC sosta task on 8 separate epochs from 
MJD=57760.03365 to 57770.50380. We used  $N_{\rm H}=4.6\times10^{20}$\,cm$^{-2}$ and a photon index of $\Gamma=1.6$ 
\citep[as in][]{GCN20415} to convert the measured counts\,s$^{-1}$ to \ergcm2s using HEASARC WebPIMMS.  A subsequent enhanced position was published on the Swift website as RA, Dec = 138.30789, 61.09263 with an error radius of 2.2 arcsecond (90\% confidence) which is also consistent with ATLAS17aeu.
The unabsorbed fluxes over the 0.3-10\,keV range are reported in Table\,\ref{tab:data}
and plotted in Fig.\,\ref{fig:lc}.
 The HEASARC tool xselect was used to extract XRT images in different energy range frames from which the photons were counted. A histogram of the energy distribution of individual photons from the 5 epochs with x-ray detection is shown in Fig.\,\ref{fig:xrt}. The values cluster around 1\,keV, consistent with a soft X-ray source. 
Two radio detections were  reported, the first by \cite{GCN20425} with the AMI-LA 
(Arcminute Microkelvin Imager Large Array)
at 15\,GHz 
on MJD=57760.04 (2.6\,days after the GW detection) 
and shortly later by 
\cite{GCN20396} with the VLA at 6\,GHz 
in an observation covering MJD=57760.58796 to 57760.62950.  
The AMI results are publicly available\footnote{https://4pisky.org/ami-grb/} and 
are listed in Table\,\ref{tab:data}. 

A gamma-ray burst  was, independently of all these data, 
discovered by POLAR \citep[GRB170105A;][]{GCN20387} with  
 a discovery time that could imply a link between the 
GRB and ATLAS17aeu
\citep[as initially proposed by][]{GCN20393}, or perhaps linking either of the events to GW170104. 
The GRB was also detected by  AstroSat CZTI \citep{GCN20389},  Konus-Wind, and INTEGRAL SPI-ACS
\citep{GCN20406}.  The properties  of the GRB are summarized in Table\,\ref{tab:data}. 
The GRB fluence from Konus-Wind was  $2.56_{-0.13}^{+0.18}\times10^{-6}$\,erg\,cm$^{-2}$,
within the energy range  20\,keV - 10\,MeV \citep{GCN20406}. The signal in the 
three energy bands 
of Konus-Wind \footnote{http://www.ioffe.ru/LEA/GRBs/GRB170105\_T22450/}
shows no flux  in the hardest band  300-1160\,keV, but significant detections
in the 18-70 and 70-300\,keV channels.  
We estimated the $T_{\rm 50}$ and $T_{\rm 90}$ durations (the time intervals containing 25-75\%  and 
5\%-90\%  of the total fluence) 
from the available Konus-Wind 
lightcurve and found $T_{\rm 50}=1.1$\,s, $T_{\rm 90}=2.8$\,s in the publicly 
available integrated 
energy bandpass of 50-200\,keV \footnote{https://gcn.gsfc.nasa.gov/konus\_grbs.html}. 
\cite{2017arXiv170600024B} reported a $T_{90}=15\pm 1$\,s
and with no signal above  100\,keV, they suggest 
a classification of a long, soft GRB.
While we measured $T_{\rm 90}=2.8$\,s in the integrated energy bandpass of 
50-200 keV (and $T_{\rm 50}=1.1$\,s), it appears that there is a flux over a longer duration of order 
20\,s in the softest channel only 
\citep[18-70keV][]{GCN20406}. This is visible in 
the online plot of the GRB duration (in the low
time resolution plot) but is not captured in the
high time resolution data files. 
\cite{2016ApJS..224...10S} presented the second Konus-Wind sample of short GRBs, and 
 in defining the sample they adopted $T_{50}<0.6$\,s. This is somewhat of an arbitrary cut off, 
 in order to define a relatively clean sample of short bursts which could be further 
 sub-classified. However \cite{2016ApJS..224...10S} more quantitatively 
 defined two Gaussian distributions in the hardness$-$duration plane for 
 1143 Konus-Wind bursts. 
The long, 20\,s duration of GRB170105A in the soft $\gamma$-ray channel 
puts it comfortably within the long GRB Gaussian distribution, and outside
the 3$\sigma$ contour that defines the short, hard GRBs. 
The hardness ratio from Konus-Wind is defined as the ratio of counts in the G3 to G2 bands ($HR_{32}$). These quantitative fluxes are not available publicly yet but since there is no flux in the 300-1160 keV range (roughly band G3) then it is likely that this GRB is long and soft. The properties of this GRB in context with the Swift, {\it Fermi} and Konus-Wind detected GRBs is further discussed in Section \ref{sec:disc}.

The detections by Konus-Wind, and INTEGRAL (SPI-ACS) allowed triangulation of the signal, resulting in a sky annulus which has a radius of 34.255$^{\circ}$  (and is 16.644$^{\circ}$ wide) centered on  RA=129.749\,deg (08h38m60s) and DEC=+27.904\,deg (+27d54m14s) as reported in \cite{GCN20406}. ATLAS17aeu is 34.040$^{\circ}$ from the center and within the annulus and therefore the positions are consistent.  
Figure\,\ref{fig:skymap} shows the skymap from LIGO for GW170104 and the likely GRB170105A 
annulus from \citep{GCN20406} together with the position of ATLAS17aeu. 
A joint AstroSat CZTI
+ IPN localization analysis showed ATLAS17aeu spatially consistent with the detection \citep{2017arXiv170600024B}. 

Therefore we have three distinct astrophysical events GW170104, ATLAS17aeu and 
GRB170105A which are spatially coincident within the uncertainties and are 
temporally coincident within 24\,hr.  We now consider the possibility that 
ATLAS17aeu is related to either of these transients and the likelihood that all three are physically related.

\begin{table*}[]
\centering
\caption{Photometry and multiwavelength fluxes 
for ATLAS17aeu and data for GW170104 and GRB170105A for comparison.  Data not presented in this paper were
released as the GCNs cited. The Swift observations and data were reported in \cite{GCN20415,GCN20390} but we re-analysed the data in the archvie to produce the numbers quoted. The AMI data are availabe publicly at the quoted website.}
\label{tab:data}
\begin{scriptsize}
\begin{tabular}{llllll}\hline 
Telescope & Magnitude/Flux/Fluence      & Filter/Waveband &  MJD  & Position    & Ref    \\     \hline 
GW170104   & ... & ... & 57757.42498 & &  \cite{gw170104} \\ \hline 
GRB170105A   & &     &    & see Fig.\,1 &  \\            
POLAR      &                           &   80-500 keV   &  57758.218137         &     &    \cite{GCN20387}\\
AstroSAT CZTI      &                           &   40-200 keV   &  57758.218125        &     &    \cite{2017arXiv170600024B}\\
Konus-Wind         & $2.56_{-0.13}^{+0.18}\times 10^{-6}$ erg\,cm$^{-2}$ & 20\,keV - 10\,MeV & 57758.218174  & Centre : 129.749,+27.904  &  \cite{GCN20406} \\
INTEGRAL-SPIACS    &                             &  80keV - 8MeV  &    57758.218125  &  Annulus radius :34.255$_{-14.832}^{+1.812}$ & \cite{GCN20387}     \\ 
\hline 
SWASP &     $>17.0$      & $r$ &57758.05948013 &  at 17aeu coords  & \cite{GCN20434}\\
SWASP &         $>16.9$  & $r$ &57758.09198558 &  ... & \cite{GCN20434}\\
SWASP  &         $>16.3$  &$r$ &57758.20992235 &  ...  & \cite{GCN20434} \\
PS1       & 17.75 $\pm$0.01   & \ips &    57758.389 & 138.30783, +61.09272 & this paper \\
ATLAS     & 18.12   $\pm$0.09            & $c$      & 57758.41297 & 138.30789 +61.09267  & this paper \\
ATLAS     & 18.25   $\pm$0.11            & $c$   & 57758.41446  & 138.30789 +61.09267 & this paper\\
ATLAS     & 18.26   $\pm$0.11             & $c$      & 57758.42672  & 138.30789 +61.0926  & this paper \\
ATLAS     & 18.57   $\pm$0.14            & $c$      & 57758.44191  & 138.30789 +61.09267  & this paper \\
ATLAS     & 18.48   $\pm$0.13            & $c$      & 57758.44691 & 138.30789 +61.09267 & this paper  \\
ATLAS     & 18.34   $\pm$0.12          & $c$      & 57758.44792 & 138.30789 +61.09267   & this paper \\
ATLAS     & 18.45   $\pm$0.10             & $c$      & 57758.45499  & 138.30789 +61.09267 & this paper  \\
ATLAS     & 18.94   $\pm$0.17            & $c$      & 57758.46196   & 138.30789 +61.09267 & this paper \\
PS1       & 18.25  $\pm$0.02  & \ips &    57758.464 & ...  & this paper \\
PS1       & 20.64 $\pm$0.10   & \ips &    57759.338 & ... & this paper \\
PS1       & 20.71 $\pm$0.13   & \ips &    57759.354 & ... & this paper \\
PS1       & 20.97 $\pm$0.09   & \ips &    57759.539 & ... & this paper \\
PS1       & 21.20 $\pm$0.12   & \ips &    57759.556 & ...& this paper \\
PS1       & 20.88 $\pm$0.10   & \ips &   57759.618  & ...& this paper \\
Swift - XRT &  $2.9\pm1.3\times10^{-13}$   \ergcm2s  & 0.3-10 keV &    57760.03365 & 138.3059 +61.0919 & \cite{GCN20390} \\
Swift - XRT &  $2.5\pm0.5\times10^{-13}$   \ergcm2s  & 0.3-10 keV &    57760.86307 &  ... & this paper \\
AMI & 336.0 $\pm20.0$ $\mu$Jy & 15.5 GHz & 57760.1667 & Not quoted in GCN & \cite{GCN20425}\\
Hale  & $21.9\pm0.3$  & $i$  & 57760.33719 & ... & \cite{GCN20393} \\
DCT       & $22.08 \pm0.05$        & $r$      &  57760.45556     & ... &    \cite{GCN20397} \\
DCT       & $21.96  \pm0.05$       & $i$      &  57760.45556     & ... &    \cite{GCN20397} \\
DCT       & $22.45  \pm0.05$        & $g$      & 57760.45556    & ... &    \cite{GCN20397} \\
VLA & 159.0 $\pm$9.8) $\mu$Jy & 6 GHz & 57760.62950 &  Not quoted in GCN & \cite{GCN20396} \\
AMI &	353$\pm$17 $\mu$Jy  &	15.5GHz &	57761.03    &  ... &  \scriptsize{https://4pisky.org/ami-grb/} \\
TNG       & $22.5 \pm0.3$    & $I$ Vega & 57761.09319  & ... &    \cite{GCN20416} \\
Gemini    & 22.77  $\pm0.17$          & $r$      & 57761.51968  &  ... & this paper   \\
Swift - XRT &  9.1$\pm$2.8$\times10^{-14}$   \ergcm2s  & 0.3-10 keV &    57762.69006 & 138.3059 +61.0919 & \cite{GCN20415}\\
Swift - XRT &  $<2.6\times10^{-13}$        \ergcm2s& 0.3-10 keV &	    57764.12551 & ...   & this paper \\
AMI &	183$\pm$19 $\mu$Jy  &	15.5GHz &	57765.03    & ...	&   \scriptsize{https://4pisky.org/ami-grb/} \\
Swift - XRT &  1.3$\pm$0.6$\times10^{-13}$   \ergcm2s  & 0.3-10 keV &    57765.38889 & ... & this paper\\
Swift - XRT &         $<2.2\times10^{-13}$   \ergcm2s& 0.3-10 keV &	    57766.21122 & ... & this paper\\
AMI &	142$\pm$19 $\mu$Jy  &	15.5GHz &	57767.02    &...	&   \scriptsize{https://4pisky.org/ami-grb/} \\
Swift - XRT &          $<1.9\times10^{-13}$  \ergcm2s& 0.3-10 keV &	    57767.37897  & ... & this paper\\
AMI &	122$\pm$20 $\mu$Jy  &	15.5GHz &	57768.03    &...	&   \scriptsize{https://4pisky.org/ami-grb/} \\
AMI &	118$\pm$27 $\mu$Jy  &	15.5GHz &	57769.03    &...	&  \scriptsize{https://4pisky.org/ami-grb/}  \\
AMI &	118$\pm$27 $\mu$Jy  &	15.5GHz &	57770.03    &...	&  \scriptsize{https://4pisky.org/ami-grb/}  \\
Swift - XRT &  5.5$\pm$2.3$\times10^{-14}$   \ergcm2s   & 0.3-10 keV &   57770.50380      & ... & this paper \\
AMI &	81$\pm$23  $\mu$Jy  &	15.5GHz &	57771.03    &...	&  \scriptsize{https://4pisky.org/ami-grb/}  \\
AMI &	149$\pm$21 $\mu$Jy  &	15.5GHz &	57772.04    &...	&  \scriptsize{https://4pisky.org/ami-grb/}  \\
AMI &	84$\pm$28  $\mu$Jy  &	15.5GHz &	57773.04    &...	&  \scriptsize{https://4pisky.org/ami-grb/}  \\
AMI &	$<81$     $\mu$Jy   &	15.5GHz &	57774.01    &...	&  \scriptsize{https://4pisky.org/ami-grb/}  \\
AMI &	$<84$     $\mu$Jy   &	15.5GHz &	57775.97    &...	&  \scriptsize{https://4pisky.org/ami-grb/}  \\
AMI &	82$\pm$18  $\mu$Jy  &	15.5GHz &	57777.01    &...	&  \scriptsize{https://4pisky.org/ami-grb/}  \\
\hline 
\end{tabular}
\end{scriptsize}
\end{table*}

\begin{figure*}
\includegraphics[width=18cm,angle=0]{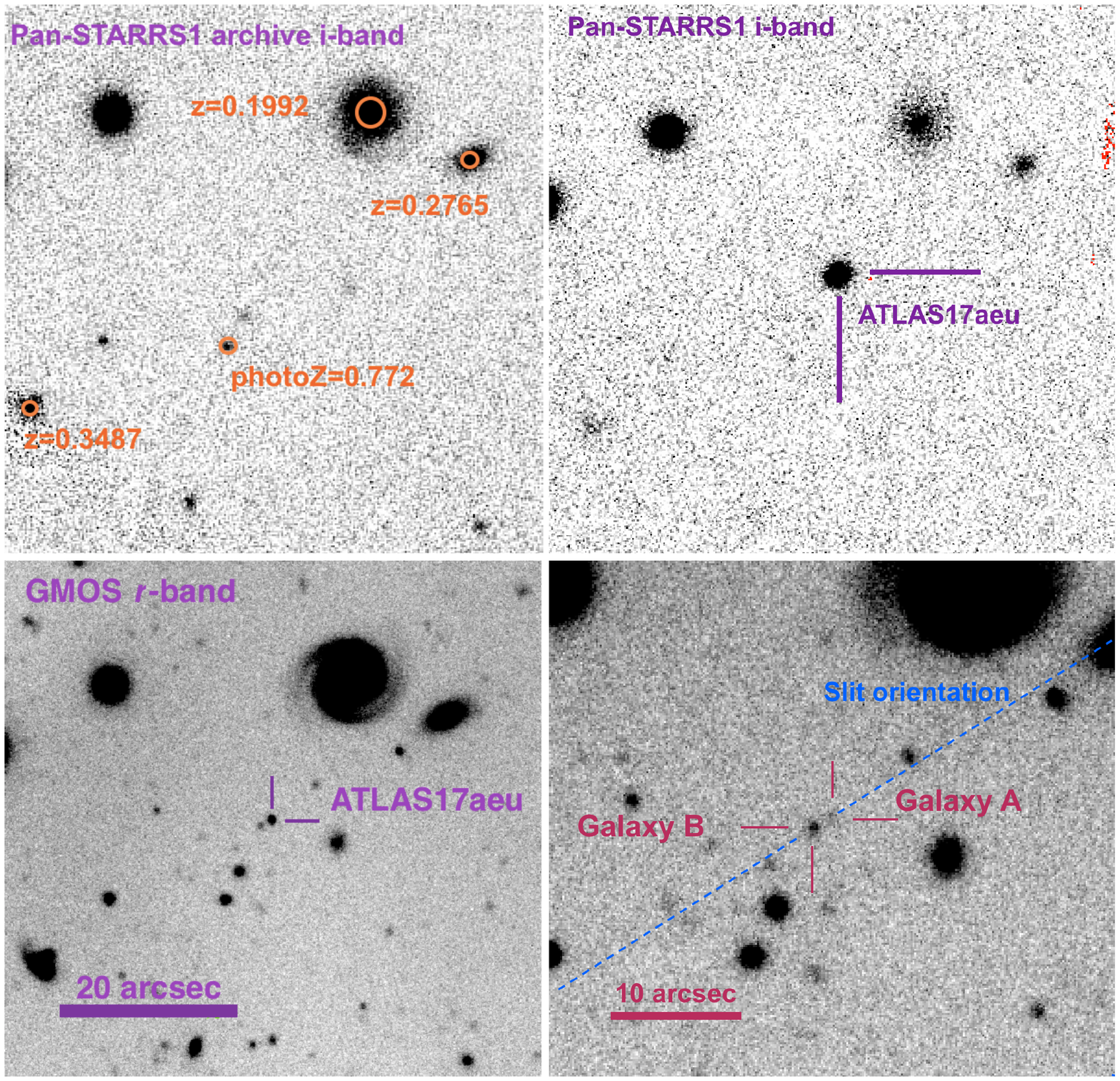}
\vspace{-2.5in}
\caption{{\bf Top Left:} The Pan-STARRS1 \ips-band archive image \citep{2016arXiv161205560C}, showing no host down to \ips$\simeq23$ at the position of ATLAS17aeu. The redshifts of neighbouring galaxies are shown 
with the $z=0.1992$  and the $z_{\rm photo}=0.772$ values from SDSS DR12 and the other two from our GMOS and Keck spectra. 
{\bf Top Right:} The Pan-STARRS1 detection of ATLAS17aeu  on 57758.389 ($\ips=17.75$)
{\bf Bottom Left:} Gemini-North GMOS image of ATLAS17aeu ($r-$band) on MJD=57761.51968, showing the object at $r=22.77$
{\bf Bottom right:} Deep Gemini-North GMOS image of ATLAS17aeu ($r-$band) at 86 days after discovery. The host
of ATLAS17aeu is Galaxy A, and the nearby brighter Galaxy B (1\farcs8 separation) is also labelled. The slit orientation
for the Gemini GMOS and Keck DEIMOS spectra is shown, each slit was 1\farcs0 wide. }
\label{fig:image}
\end{figure*}

\section{Host galaxy and redshift constraints for ATLAS17aeu}
\label{sec:host}
The relative volumetric rates of GRBs, GRB-like afterglows and GW sources can inform
a discussion of probability of coincidences. Therefore identification of the host
galaxy and redshift determination of ATLAS17aeu is desirable. The initial 
GMOS spectrum of ATLAS17aeu, when it was fading at $r=22.77$, shows a blue 
featureless continuum (Fig.\,\ref{fig:lc}). There are no obvious absorption lines
such as Mg\,{\sc ii} (from ISM absorption in the host) and no strong 
nebular emission lines (e.g. H$\alpha$, O\,[{\sc iii}]) from star formation in the host. 
The blue featureless continuum can be fit with a power-law with index $\alpha=-1.92$. This is not unlike other GRB afterglow spectra,
for example from the extensive study of \cite{2009ApJS..185..526F}. Some of these are examples for
which a definitive redshift has proven difficult. Two examples are 
shown in (Fig.\,\ref{fig:lc}), GRB060110A and GRB080523 from  \cite{2009ApJS..185..526F}. 
The shape of the SED of ATLAS17aeu from the Gemini spectrum (which is +3.3 days after GRB170105A) is consistent with the color inferred from the 
ATLAS cyan and 
the Pan-STARRS $\ips$ photometry at just +4.8\,hr after the
GRB. The SED from the Discovery Channel Telescope 
photometry  (at +2.3\,days after GRB170105A) by \cite{GCN20397} also shows the same 
blue slope. This indicates that the optical SED of ATLAS17aeu stayed
relatively constant for the three days over which it was observed with a power-law  index $\alpha=-1.92$.

A montage of our images is shown in Fig.\,\ref{fig:image}.  We originally noted in \cite{GCN20382} that ATLAS17aeu  is
23 arcsec from the face-on, bright, spiral galaxy SDSS J091312.36+610554.2 which
has a spectroscopic redshift of $z = 0.19900\pm0.00004$.  This implies a luminosity distance of 990 Mpc so if ATLAS17aeu were related it would be at a projected distance of 75\,kpc from this galaxy (see Fig.\,\ref{fig:image}).
This luminosity distance is consistent with the LIGO distance 
to  GW170104 which suggests a 10-90 percentile probability of 520-1010\,Mpc in this direction
\citep[from the analysis of][]{2016ApJ...829L..15S}. 
However our deep GMOS imaging reveals the environment of ATLAS17aeu in more detail. 
There is a faint galaxy exactly at the position of ATLAS17aeu in the 
Gemini image of 2017 Apr 01, which we label "Galaxy A" in Fig.\,\ref{fig:image}. 
With aperture photometry, we measure $r=25.59\pm0.16$. There is also a
brighter extended source which is 1\farcs8 south east of the position of 
ATLAS17aeu  (RA=9:13:14.120, DEC=+61:05:32.60), for which we measure 
$r=24.44\pm0.09$. We label this "Galaxy B", and it is clearly visible in the 
earlier Gemini image (from 2017 Jan 08) resolved from ATLAS17aeu. In this image
we measure $r=24.11\pm.09$, with the difference between the two arising from the 
choice of aperture required to minimise contamination from the bright flux of 
ATLAS17aeu. Galaxy B is almost certainly the object report by 
\cite{GCN20735} at $r=24.23\pm0.2$, which they proposed as a host. 
Our GMOS images now show that this is offset and resolved from the position of 
ATLAS17aeu. It is possible that Galaxy A and B are 
physically linked and could either be a merging or 
disturbed system and a redshift for both is desirable.

\begin{figure*}
\centering
\includegraphics[width=18cm,angle=0]{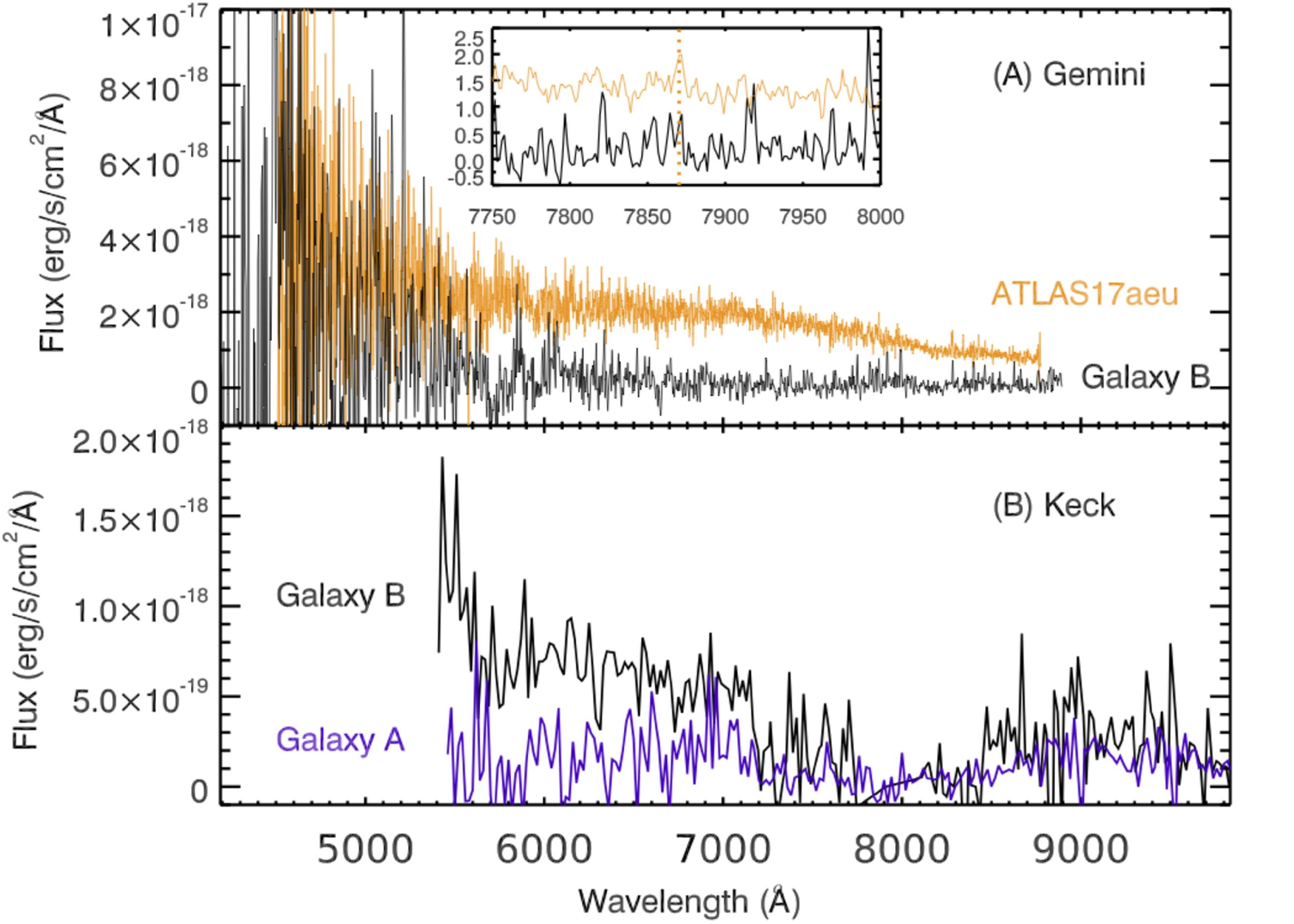}
\caption{{\bf Top :} The 2D spectral image from Gemini + GMOS taken when 
ATLAS17aeu had faded, in order to search for any residual host galaxy emission lines. The faint trace of Galaxy B is labeled, and there is no discernible 
flux at the position of Galaxy A. 
{\bf Bottom:} Extracted, flux calibrated spectra from GMOS and Keck. In panel (A) the ATLAS17aeu spectrum 
from Gemini on 57761 (3.3\,days after GRB170105A) is in orange. The later, deep spectrum of extraction of Galaxy B (shown in the 2D image) is in black. Galaxy A is not clearly visible in the GMOS spectra, although a forced extraction can be done. The possible emission line in
the ATLAS17aeu spectrum at 7870.3\,\AA, is not clearly visible above the noise in the deeper
2D spectral image or in the extracted spectrum, as
indicated in the inset.
 In Panel (B) the extracted Keck DEIMOS spectra of Galaxy A and B are shown. The skylines dominate the red and were manually snipped out, after which the spectra were rebinned to 20\,\AA\ per pixel dispersion. Synthetic photometry was used to scale the flux level to match the measured $r$-band magnitudes in the difference image. }
\label{fig:spec2D}
\end{figure*}

We placed the slit of GMOS and DEIMOS (see Sect.\ref{sec:obs} for details of these deep spectra) across
these two galaxies as illustrated in Fig.\,\ref{fig:image} in search 
of any emission lines of either which would provide a redshift for 
ATLAS17aeu. There is no obvious emission line detected at the spatial 
positions of Galaxy A or B. The slit passed through two brighter galaxies, for which we measure 
redshifts $z=0.3487$ for the starforming SDSS J091318.49+610512.6
and 
$z=0.2765$
for the early type galaxy SDSS J091310.39+610548.6
(labeled in Fig.\,\ref{fig:image} for completeness).
In the Gemini GMOS spectrum (for which nod and shuffle was used),    there is faint continuum present for Galaxy B, whereas no apparent signal is 
at the position of Galaxy A (see Fig.\,\ref{fig:spec2D}). 
 We are confident in the location of Galaxy B on the 
2D spectral images from the spatial consistency of its position and measured offsets
from the $z=0.3487$ and  $z=0.2765$
bright sources located in the slit.  In the first GMOS spectrum of
ATLAS17aeu there is weak, possible emission line at 7860.3$\pm$0.5\,\AA\ 
illustrated in the inset panel of  Fig.\,\ref{fig:spec2D}. 
If this was confirmed, it could potentially be H$\alpha$ from Galaxy A or B at $z=0.199$. The feature is visible in two different extraction methods, 
either extracting and combining the six individual spectra or combining and 
sky subtracting the 2D frames and then extracting the object. It has 
a FWHM of $5.7\pm0.7$\,\AA, which is narrower than the skylines measured on the
frames (6 to 7.6\,\AA).  No spatially extended 
flux is  visible on the 2D spectral frame at this position.  
An excess in flux is only marginally visible in the 2D 
spectrum of ATLAS17aeu at this position (which would be 
Galaxy A). 
However we find no confirmation of the line 
in the GMOS or DEIMOS deep spectra taken later (Fig.\,\ref{fig:spec2D}) and therefore
cannot confirm a redshift for either Galaxy A or B.
A faint continuum for each of Galaxy A and B is visible in the Keck 2D images and we forced extraction of the signal using the trace of the bright $z=0.2765$ galaxy. Since slit nodding or shifting was not used in the Keck spectra, the red part is skyline noise dominated. We manually snipped out the lines and rebinned the 1D, flux calibrated, spectra to 20\,\AA\ per pixel dispersion. This gives a course spectrum for which the overall SED should be reliable. We then applied synthetic photometry and scaling to bring the continuum into line with $r=24.1$ and $r=25.6$ respectively for Galaxies B and A.    
Fig.\,\ref{fig:spec2D} shows that Galaxy B is bluer than A, however no reliable redshift could be determined. Galaxy A looks relatively flat in its SED, while Galaxy B clearly does have a rising blue continuum (which we also see in the Gemini spectrum). 
There is no clear sign of the 4000\,\AA\ break, for example, in either. This could mean that
$z\lesssim0.25$, but it is not a definitive statement given the signal-to-noise and wavelength coverage of the data. We may be seeing either the rising optical continuum of a low redshift galaxy, or the rising 
UV (2000-3000\,\AA) continuum of an Sb-Sd starforming
galaxy. 
The fact that we see no strong absorption due to the Lyman-$\alpha$ forest in the 
first GMOS spectrum of ATLAS17aeu provides a robust upper limit to the 
redshift of $z\lesssim2.9$ \citep[as done for other GRBs in][]{2009ApJS..185..526F}. As far as we can tell with these data, there is no clear evidence of either being high redshift ($z\gtrsim1$). 
At the position of ATLAS17aeu (Galaxy A, where we see no continuum flux in the Gemini spectrum) 
 the 3$\sigma$ upper limit for 
an emission line is $3 - 4 \times 10^{-18}$erg\,s\,cm$^{-2}$ depending 
on where the measurement is made. In the Gemini spectrum of Galaxy B, the limit is
around $1\times 10^{-17}$erg\,s\,cm$^{-2}$. 
We chose to make these measurements on the Gemini spectrum only due to the superior sky subtraction methods employed during observing and reductions. 

Therefore we are left with the conclusion that ATLAS17aeu has a 
very faint host galaxy, at $r=25.59\pm0.16$ which is offset by 1\farcs8 
from another galaxy at $r=24.27\pm0.09$. We have no firm redshift constraints
for either, nor can we say if they are physically (kinematically) linked. 
 The sample of host galaxies of 46 GRBs from 
\cite{2009ApJ...691..182S}, which extends out to redshift $z\simeq2$, indicates that about
13\% (6 out of 46) are located in galaxies $r=25$ or fainter. 
\cite{2015A&A...581A.125K} 
shows that for GRBs with $z\lesssim0.5$, 
there are no objects with host galaxies fainter than 
$r\sim25$, but that they are relatively common beyond 
$z\gtrsim1$.  
If ATLAS17aeu were at the redshift of GW170104 
($z = 0.18^{+0.08}_{-0.07}$ or 
$D_{\rm L} = 880^{+460}_{-360}$\,Mpc)
then it would be only $g\simeq -14.2$ mag, 
which is very faint even for GRB host galaxies. 
Assuming this redshift, it  would imply a limit on the 
$H\alpha$ luminosity 
$L_{H\alpha} \leq 4.0^{+5.4}_{-2.6} \times 10^{38}$\,erg\,s$^{-1}$
and a limit on the SFR of 
$SFR \leq 0.003^{+0.004}_{-0.002}$\,\msun\,yr$^{-1}$. 
Of course, we have no firm redshift constraints other than 
it is $z\leq2.9$, which would correspond to 
$L_{H\alpha} \leq 3.2\times 10^{41}$erg\,s$^{-1}$
or a SFR limit of 
$SFR \leq 2.5$\msun\,yr$^{-1}$. 
In conclusion, we do not have a measured redshift
for the host of ATLAS17aeu but the faintness of the 
detected galaxy and the deep limits on H$\alpha$ emission
imply that a high redshift GRB origin ($z\gtrsim1$) for 
ATLAS17aeu would be compatible with the characteristics 
of the known population of long, soft GRBs. 
Of course if ATLAS17aeu were associated physically 
with GW170104 then differences between it and the 
known GRB population might be expected.

\section{Discussion and Analysis}
\label{sec:disc}

We begin by investigating the scenario whereby  ATLAS17aeu is indeed the optical, x-ray and radio afterglow 
of GRB170105A.   The x-ray source and the radio source are to a very high degree of probability the counterparts
of ATLAS17aeu.  The detection times of the three transients are listed in Table\,\ref{tab:data}  in decimal 
MJDs and given here in UT times to the nearest second:
GW170104 at 2017-01-04 10:11:58 
\citep{gw170104}, 
GRB170105A at 2017-01-05 06:14:07
\citep{GCN20387,GCN20389},
and ATLAS17aeu at 2017-01-05 09:54:40.
The Swift x-ray transient is within 4\farcs1 (within the 1$\sigma$ error bars) of ATLAS17aeu.  
Although the radio detections do not state their positional uncertainty, the radio instruments have pointing uncertainties of about 5\farcs0 indicating their likely positional coincidence.
The probability that either the radio or the x-ray transient is coincident with ATLAS17aeu to with 5\farcs0 is $\approx 10^{-10}$. Trivially, we must assume that these transients are from the same object, ATLAS17aeu. 

Fig\,\ref{fig:lc}  shows the optical and x-ray lightcurves of ATLAS17aeu in comparison with other 
GRB afterglows. The decline rate in the optical is a broken power law with indexes $\alpha_1=1.20 \pm 0.02$ and $\alpha_2=0.82 \pm 0.02$, which is 
compatible with other known GRB afterglows from \cite{2009ApJ...693.1484C}, and may offer insight into the circum-burst medium. The x-ray lightcurve for Swift XRT is shown in 
comparison with 297 GRBs with known redshifts and again, it sits  within the expected locus of 
points. GRBs show a trend between x-ray flux and $\gamma$-ray fluence, albeit with a broad spread of more than 
one order of magnitude \cite[see Fig.\,12 of][]{2009MNRAS.397.1177E}. 
The fluence of GRB170105A \citep[$2.56_{-0.13}^{+0.18}\times 10^{-6}$\,erg\,cm$^{-2}$;][]{GCN20406} 
should be compared with the estimate of x-ray flux at 11hrs after burst.  Fig.\,\ref{fig:lc}  indicates that this 
would be of order $10^{-12}$\,erg\,cm$^{-2}$\,s$^{-1}$ if we extrapolate back to $\sim4\times10^4$\,s after the
GRB. The  ratio of x-ray flux to $\gamma$-ray fluence thus comfortably sits in the observed range of  GRB 
afterglows.   Long GRBs show a characteristic range of optical to x-ray flux, when both are
 estimated at $t=1000$\,s  from the burst \citep{2009ApJ...693.1484C}. For ATLAS17aeu we estimate 
$F_{\rm opt} = 660\mu$\,Jy and  $F_{\rm x} =11\mu$\,Jy, having corrected the $r-band$ extrapolated point
for Galactic ISM extinction and the extrapolated XRT data for Galactic H\,{\sc I}  column density (and assuming an energy band of 1\,keV, see Fig.\,\ref{fig:xrt}). This falls within the broad range of the \citet{2009ApJ...693.1484C} sample and within the
spectral index  $0.5<\beta<1.25$ expected for GRB afterglows. 
While these numbers are compatible, we need to consider 
the rates of GRB more quantitatively to determine the probability of association of ATLAS17aeu 
with GRB170105A. 

We now calculate the  probability that the $\gamma$-ray burst is independent of ATLAS17aeu (i.e.  just a chance coincidence). We use Poisson statistics, where the probability of an occurrence of $n$ events is given by 

\begin{equation}
P(n) = \frac{e^{-\lambda}\lambda^{n}}{n!}
\label{eqn:poiss}
\end{equation}

where $\lambda$ is the expectation value. The value of 
$\lambda$ is the product of a number of factors given by the rate of each

\begin{equation}
\lambda = \prod_{i=1}^{k} r_i
\end{equation}

The rates we will discuss in this section are listed for reference in Table\,\ref{tab:rates}. 
In this case, we will derive the 
Poisson probabilities of getting one or more random coincidences, which simplifies Eqn. \ref{eqn:poiss} to 

\begin{equation}
p=1-e^{-\lambda} = 1 - e^{-\prod_{i=1}^{k} r_i}
\label{eqn:poiss0}
\end{equation}

To determine the time coincidence, we take as a lower bound the SWASP observation at MJD=57758.20992235,
\citep{GCN20434}
 and as an upper bound the first PS1 observation (to an approximate depth of 21.5 mag) at 57758.38892. GRB170105A occurred at 57758.218137.
Therefore we have a time coincidence window of ($-0.01,0.17$) days. 
The 2nd {\it Fermi} GBM catalog \citep{2014ApJS..211...13V}  and the Swift catalogue
\footnote{http://www.grbcatalog.org/}
\citep{2009MNRAS.397.1177E} indicate rates of 238 and 99 GRBs per year in each facility. However there is 
obviously overlap between the two samples.  The 
{\it Fermi} GBM catalogue indicates that 26\% of their sources were also detected by Swift, therefore we estimate 
a total sky rate of {\it Fermi} GBM + Swift of 275 per year. 
For a GRB rate of 
275 per year (0.75 per day; $r_1 = 0.75$) and a time window of 0.18 days ($r_2$=0.18), this results in a probability of getting 1 or more random coincidences of $p=0.13$. 

The positional coincidence from the IPN triangulation of 
\cite{GCN20406} means we should use the positional annulus for the GRB170105A 
which covers 2901.5 square degrees, resulting in a probability of 0.07 compared to all sky (i.e. $r_3$=0.07). 
Combining these probabilities, we get 1\% probability that GRB170105A was 
simply a chance coincidence
with ATLAS17aeu. 
($p=1 - e^{r_1 r_2 r_3}=0.01$). 
In other words the chance coincidence of ATLAS17aeu and GRB170105A can 
be ruled out with 99\% (2.6$\sigma$) confidence. 

\begin{figure}
\includegraphics[width=\columnwidth,angle=0]{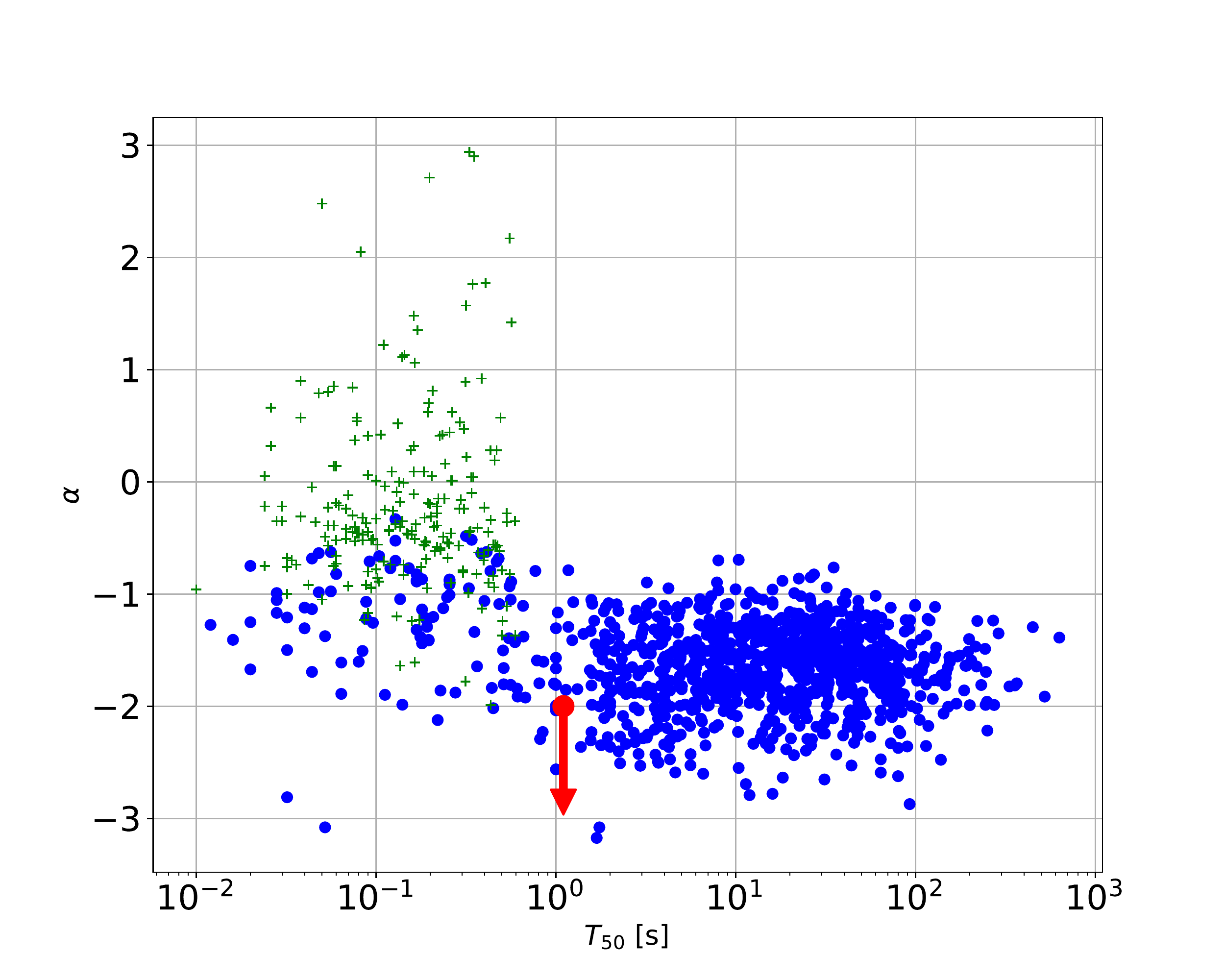}
\includegraphics[width=\columnwidth,angle=0]{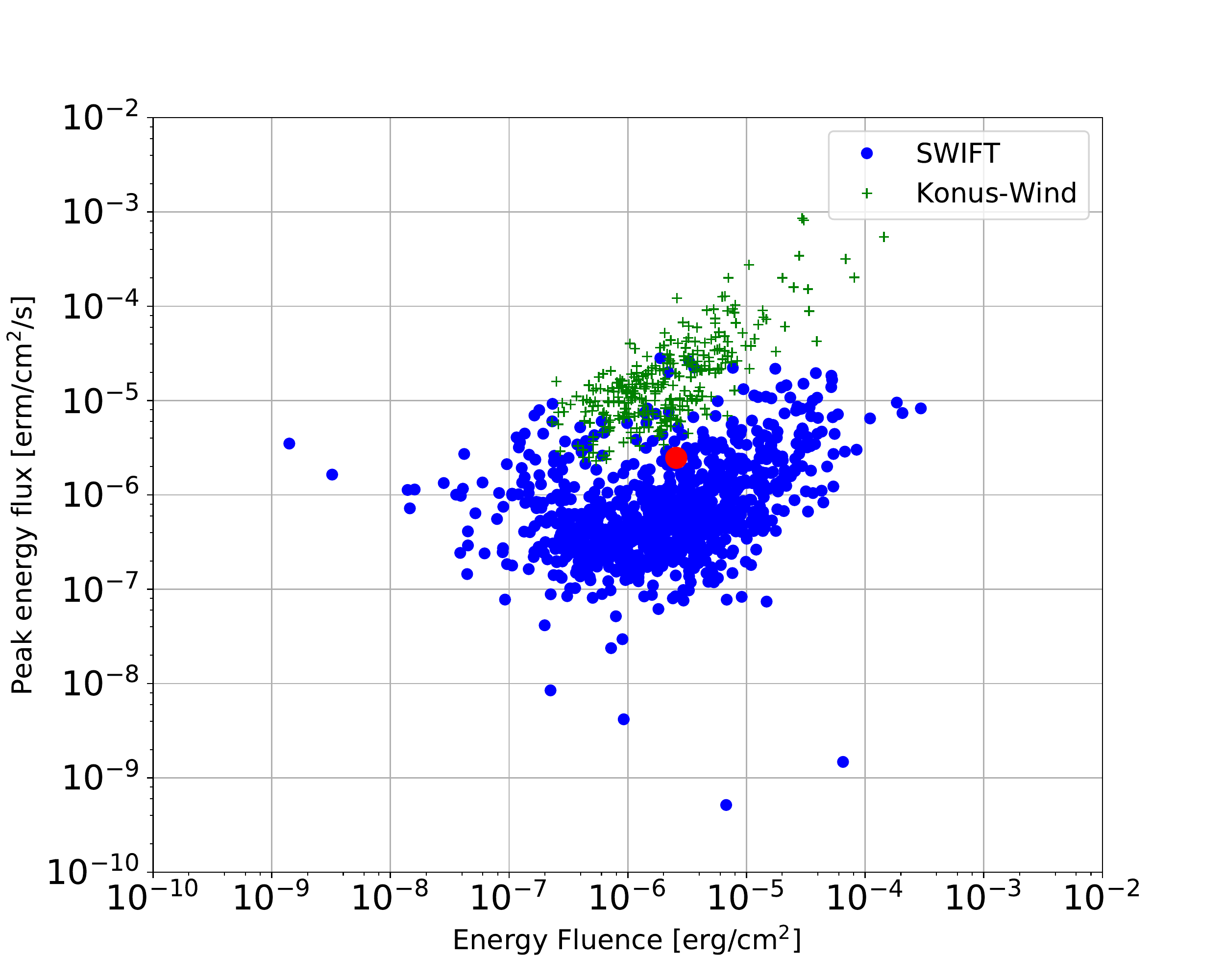}
\includegraphics[width=\columnwidth,angle=0]{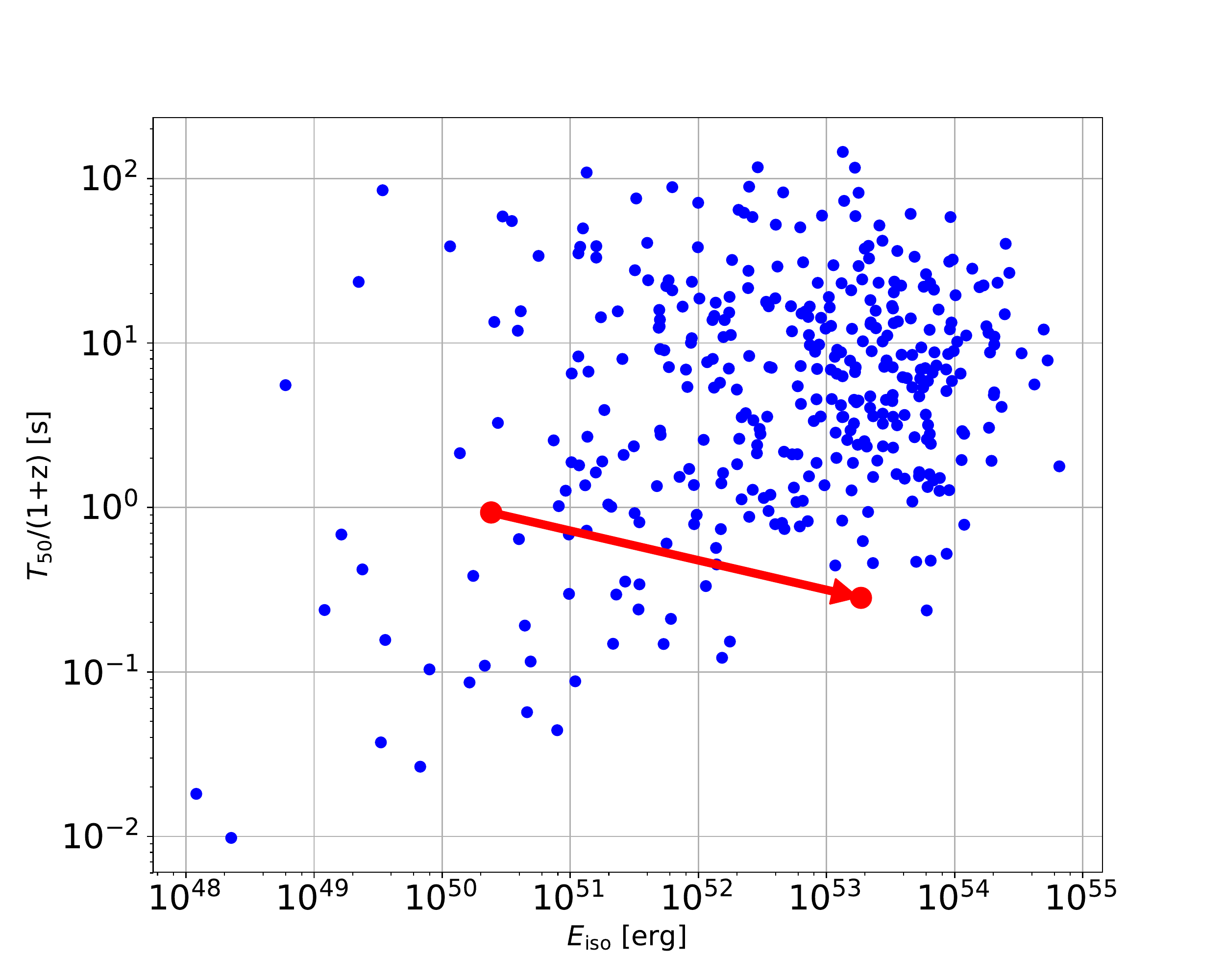}
\caption{
{\bf A (Top):} $\alpha$ vs. $T_{50}$ for Konus-Wind (short GRBs only), and SWIFT GRBs, with approximate position of GRB170105A.
An upper limit only is possible, since there is no flux in the 
highest energy band and not all data are yet available.  
{\bf B (Middle) :} Peak energy vs Energy fluence for Konus-Wind (short GRBs only), SWIFT, and GRB170105A. 
{\bf C (Bottom) :} $E_\mathrm{iso}$ vs $T_{50}$/(1+z)  for SWIFT and GRB170105A, where z has been chosen to span from z = 0.18 (median redshift for GW170104) to z = 2.9 (with the arrow pointing to higher redshift). 
}

\label{fig:grbs}
\end{figure}

This calculation assumes, of course, that ATLAS17aeu has properties that
lead us to hypothesise that an association is plausible. 
We are justified in this since we find 
that the   x-ray, optical,  and radio detections are similar to GRB afterglows. 
The three energy bands of Konus-Wind show  no detected flux in the hardest channel of 300-1160keV. This, combined with the  $T_{\rm 90}$ measurements of POLAR and AstroSat CZTI of between 2-15\,s, and the soft extended $\gamma$-ray flux  (around 20\,sec duration)
detected by Konus-Wind (18-70\,keV) would classify GRB170105A as a 
very soft GRB (as discussed above in Section\,\ref{sec:multiw}) with extended emission. 
In Fig.\,\ref{fig:grbs} we 
place GRB170105A in the locus of the Swift and Konus-Wind bursts with the measured 
$T_{50}\sim 1.1$ and with an approximate upper limit of $\alpha < -2$.  While we are aware that $T_{50}$ is larger in the 
softest channel (only), we do not have access to those
data to make a quantitative measurement and therefore
we plot the specific measurement we can make on the 
integrated 50-200\,keV data. 
Its location is not particularly unusual in the broad population of GRBs, 
although how soft the spectrum is remains to be reported by the 
Konus-Wind team. Since the redshift is unknown, the arrow in the 
$T_{50}/(1+z)$ vs $E_{\rm iso}$ plane represents the possible position of 
ATLAS17aeu from z=0.18 (were it to be associated with GW170104) and
$z=2.9$, which is our upper limit from the lack of Lyman-$\alpha$ absorption
in the Gemini afterglow spectrum. 

\cite{2014MNRAS.440.2059A} reported radio observations of the GRB130427A with the  Arcminute Mircrokelvin Imager (AMI) at a frequency of
15.7 GHz between 0.36 and 59.32 days. At a redshift of $z=0.340$, 
it had a flux of 1300$\mu$Jy at +2 days, which is 4 times brighter
than ATLAS17aeu at the same epoch (with respect to the GRB in each case)
as shown in Fig.\,\ref{fig:lc}. 
\cite{2014MNRAS.440.2059A} compiled early radio observations of 
GRBs and illustrated that the brightness temperatures can be 
estimated from 

\begin{equation}
T_{\rm b} = 1.153\times10^{-8}d^{2}F_{\nu}\nu^{-2}t^{-2}(1+z)^{-1}
\end{equation}

where $d$ is the distance (in cm), $F_\nu$  is the flux (Jy), $\nu$ is the observed frequency in Hz, and $t$ is time (s) 
since the GRB. \cite{2014MNRAS.440.2059A}  further 
discuss that a minimum Lorentz factor of a jet (assuming
the standard GRB scenario) can be estimated 
from $T_{b}/T_{B}=\Gamma^{3}$, where 
the maximum brightness temperature is the inverse-Compton limit
$T_{B} \sim 10^{12}$\,K   \citep[as in][]{1999Natur.398..394G}. 
For ATLAS17aeu to have a comfortably high Lorentz factor of 
$\Gamma>5$ \citep[the minimum found in][]{2014MNRAS.440.2059A}
then the redshift must be $z\gtrsim1$. 
Therefore, one explanation is that ATLAS17aeu is the x-ray, optical and 
radio afterglow of GRB170105A, and that the GRB is part of the 
known population of long, soft GRBs between redshift 
$1 \lesssim z \lesssim 2.9$. All the data we have
would be consistent with that explanation, including the 
low probability of a chance coincidence  of the two ($p=0.01$ as 
discussed above). 

\begin{table*}
\centering
\caption{Sky rates of GRBs and their x-ray and optical afterglows. The top three are all sky rates per year. The bottom 5 are relative fractions. }
\label{tab:rates}
\begin{tabular}{llll}\hline 
Object                  & Rate   & Reference \\\hline 
{\it Fermi} GRBs (GBM) &   238 per year  & \cite{2014ApJS..211...13V}  \\
Swift GRBs         &   99 per year   & http://www.grbcatalog.org/ \\
{\it Fermi} + Swift      &   275 per year  &    see Section 4     \\
Fraction of Swift GRBs with x-ray afterglow   & 0.95    & \cite{2009MNRAS.397.1177E} \\
Fraction of Swift GRBs with $m<17$ optical afterglow   &   0.3 & 
 \cite{2009ApJ...693.1484C} \\
Fraction of GRBs in $r\geq25$ hosts & 0.13 & \cite{2009ApJ...691..182S} \\
Fraction of LGRBs within $z<0.25$ & 0.025 & \cite{2014ARAA..52...43B} \\
Fraction of all (short+long) sGRBs within $z<0.25$ & 0.158 & 
\cite{2014ARAA..52...43B} \\
\hline 
\end{tabular}
\end{table*}

We assume then that  ATLAS17aeu is the optical, x-ray and radio afterglow of GRB170105A and  now consider the rate of GRBs to determine the Poisson probability 
of coincidence of this gamma-ray, x-ray, optical and radio transient 
 with GW170104. 
The GRB170105A was approximately 
$t$=1 day away from the GW170104, hence the Poisson probability of 1 or more GRBs per 1 day period is 
$p=1- e^{-r_1}=0.53$ (where $r_1=0.75$).
The LIGO skymap has an area of 2000 square degrees  
\citep[90\% enclosed probability, from the LALInference map;][]{2015PhRvD..91d2003V} or about 0.05 of the sky. 
Therefore the rate of GRBs within 1 day and 2000 square degrees provides a 
probability of coincidence of (for $r_2$=0.05) 
$p=1- e^{-r_1 r_2}=0.04$.  This is the probability in 3 dimensions - the 2D location on the sky and in the temporal 
window of 1 day.  In other words 
a chance coincidence is unlikely at 2.1$\sigma$ level, but not ruled out. 

However it is worth considering the facts that
if GRB170105A is associated with ATLAS17aeu then 
there are other properties that make the chance of 
coincidence  reduce.  The rate of Swift GRBs which have $m<17$ optical afterglows is 30\% of all Swift GRBs
from \cite{2009ApJ...693.1484C}. We chose this value of  $m<17$ since \cite{2009ApJ...693.1484C} 
defined their afterglow distribution at $t=1000$\,s after the GRB and extrapolating ATLAS17aeu back to 
$t=1000$\,s, it would comfortably sit above this flux level. With only 30\% of GRBs having bright 
afterglows, this reduces the probability of an optically bright GRB within 24hrs, within
the sky annulus to 
$p=1- e^{-r_1 r_2 r_3}$=0.01 (with $r_3=0.3$)
In other words a chance coincidence is unlikely at the 2.6$\sigma$  level.  
One could also propose that the host galaxy of ATLAS17aeu (Galaxy A) is
unusually faint for a GRB host, in that \cite{2009ApJ...691..182S} find 
that only 13\% of GRBs are found in galaxies fainter than $r=25$. Hence
if we fold this into the Poisson probability equation, we reduce the 
expectation value by a factor of $r_4=0.13$, giving 
$p=1- e^{-r_1 r_2 r_3 r_4}$=0.001
This would imply that 
a chance coincidence is unlikely at the 99.9\% level, or 3.3$\sigma$.  However we recognize that inclusion of every
property of GRB170105A simply serves to reduce the probability and therefore reduces the meaning and viability of the calculation. We therefore simply state that the chance of 
a long GRB within the time and 2D sky area window 
for GW170104 is 
$p=0.04$. While we have a 4D location (in space and time)
for GW170104, we lack a definitive distance to GRB170105A.
Such a measurement would provide a more definitive 
answer. 

While the probability of a chance coincidence of the GW even with GRB170105A
(which has ATLAS17aeu as its afterglow) is low, there 
are further reasons 
to be cautious about drawing a conclusion of a causal link. 
We lack a quantitative 
physical mechanisms to  
produce a GRB and multi-wavelength afterglow beginning 24hrs after the 
merger of two black holes of masses 31\msun and 19\msun. \cite{2016ApJ...821L..18P} and \cite{2017ApJ...839L...7D} have proposed that binary black holes 
could well have a circumbinary disc still existing at the time of merger and 
that perturbation of this disk by the GW energy could result in accretion onto 
the newly formed single BH with energy from the mid-energy x-ray range to the
mid-infrared.

For illustrative purposes, to consider the plausibility that emission of 
the sort proposed by \cite{2016ApJ...821L..18P} and \cite{2017ApJ...839L...7D} has occurred, we 
speculate that ATLAS17aeu is within the distance range implied by 
LIGO ($z = 0.18^{+0.08}_{-0.07}$ or $D_{\rm L} = 880^{+460}_{-360}$\,Mpc). 
We consider the x-ray and optical to be linked through 
emission from hot, thermal radiation from a circumbinary disk. 
The measured contemporaneous fluxes at $+3$days are
$\nu f_{\nu}=3.7\times10^{-14}$\ergcm2s in the optical (\ips=21.5)
and 
$\nu f_{\nu}=2.9\times10^{-13}$\ergcm2s in the x-ray.
This gives a ratio of 
$\nu f_{\nu}^{\rm x-ray}/\nu f_{\nu}^{\rm optical}= 7.8$. 
The Swift detected x-rays were soft as shown in Fig.\,\ref{fig:xrt}, 
peaking at 1keV or below. 
If we assume that they peak at 300 eV, 
we can approximately fit a thermal blackbody SED 
at $T_{\rm BB} \sim 200,000$\,K (20eV), which would peak in the EUV around 140\AA. 
However a thermal blackbody 
spectrum is a poor fit to the Gemini afterglow spectrum and the 
inferred SED from the combined ATLAS and PS1 colors and the DCT 
photometry of \cite{GCN20397}, and also does not explain the radio flux. 
Therefore we can conclude that the spectral 
energy distribution at the time of the first (and brightest) x-ray 
point appears to be non-thermal, no matter what distance the GRB
is at.  If  indeed ATLAS17aeu was within the LIGO distance 
range then it has luminosities in the x-ray, optical and 
radio ($\nu L_\nu$ in Table\,\ref{tab:energy}) of order $\sim10^{42}-10^{43}$\,erg\,s$^{-1}$. 
This is broadly similar to the energies predicted by \cite{2017ApJ...839L...7D}
for the circumbinary disk accretion model. The uncertainties are large 
and the luminosity is a strong power of $v$, the greater of the 
Keplerian disk velocity and the post-merger kick velocity (which 
is unknown), and is dependent on an unknown efficiency scaling factor and the 
relic disk mass (see their equation 6). 
In addition, the 
de Mink - King model predicts that the emitting region should have a 
characteristic emitting temperature between $10^{6}\,K \lesssim T\lesssim 10^{7}\,K$
and the peak of the EM radiation should be in the medium x-ray energy 
regime. This is quite compatible with the emitted 
energies of ATLAS17aeu (Table\,\ref{tab:energy}) and the delay
time of $\sim$hrs between the GW and energy emission is also in the 
broad region proposed in \cite{2017ApJ...839L...7D}. 
Of course, this 
relies on the assumption that ATLAS17aeu is within the LIGO distance 
range. It's not clear if the de Mink - King model can also produce a GRB, 
through disk accretion 
with a delay of $\sim$24\,hrs after black hole merger. 
It is possible that the GRB emission is a distraction, a chance coincidence,
and not related to ATLAS17aeu. As shown above this is statistically 
unlikely but still plausible. 
A key requirement is future constraints on the redshift of the host galaxy of 
ATLAS17aeu, either through very deep and long integration spectra or  through a photometric redshift technique. 
If we were able to determine the redshift and luminosity distance of either Galaxy A or B, and
if it were to fall within the distance range of GW170104, this would reduce the rate of such a coincidence to levels that would force consideration of a physical link. 
Correspondingly, if the redshift were definitive in being outside the GW170104 range, a chance coincidence could be securely concluded.

\begin{table}
\centering
\caption{Luminosities and energies at different wavelengths. The fluxes 
are from the sources reported in the text and the $\nu L_\nu$ assume, 
for illustrative purposes, the luminosity distance of  GW170104,  
$D_{\rm L} = 880$\,Mpc. The gamma rays are measured in 
the band 20\,keV - 10\,MeV, and the x-rays in the Swift band 0.3-10 keV.}
\label{tab:energy}
\begin{tabular}{lllll}\hline 
Wavelength  & Epoch    & $\nu f_\nu$   &   $\nu L_\nu$ &  $E_{\rm iso}$\\\hline 
            &          & erg/s/cm$^{2}$ &   erg/s      &   ergs \\\hline 
Gamma rays  & 57758.22 &     ...        & ...         &  2.4$\times10^{50}$ \\ 
Optical     & 57758.39 &     1.2$\times10^{-12}$ &   1.0$\times10^{44}$  &  ...       \\ 
X-rays      & 57760.03 &     2.9$\times10^{-13}$ &   2.7$\times10^{43}$  &  ...       \\ 
Optical     & 57760.03 &     3.7$\times10^{-14}$  &  3.4$\times10^{42}$  &   ...      \\
Radio       & 57760.17 &     5.0$\times10^{-16}$  &  4.7$\times10^{40}$  &  ...       \\
\hline 
\end{tabular}
\end{table}

Finally, we comment on how frequently ATLAS should see the optical afterglows of GRBs during its normal operations. Since ATLAS typically observes 
5 times over a 5000 square degree footprint every night, there is a 
quantifiable probability of 
the system catching a GRB afterglow. The number 
$N_{\rm GRB}$ afterglows that ATLAS is likely to detect per year is 

\begin{equation}
N_{\rm GRB} =  3.2 \frac{R_{\rm GRB}}{365}\frac{A}{5000}\frac{w}{5}\frac{f_{\rm obs}}{0.5}\frac{C}{0.7}\,\, {\rm yr^{-1}}
\end{equation}

where $R_{\rm GRB}$ is the annual rate of GRBs, $A$ is the area in square degrees covered per night, $w$ is the time window in hours that GRB afterglows are typically visible above the ATLAS limit of $c, o \simeq 18.5$, $f_{\rm obs}$ 
is the fraction of GRB afterglows that are detectable by ATLAS during this 
period of $w$\,hrs and $C$ is the fraction of clear, useful weather time that ATLAS 
observes. For $R_{\rm GRB}=275$, then $N=2.4$ over one year 
with the major uncertainty being
$f_{\rm obs}$. Since ATLAS has been working in survey mode for approximately
one year (March 2016 - April 2017) and we have been processing the difference imaging 
routinely during this period, finding 1 candidate for a GRB-like afterglow
(ATLAS17aeu) is not unexpected ($p=30\%$ of finding 1 or 0 for an expected rate of 2.4). 

\section{Summary}
We have reported the discovery of the transient ATLAS17aeu 
which lies spatially within the GW170104 skymap 
and has a first optical detection 23.2hrs after the binary black hole 
merger. Analysis of 
multi-wavelength data from x-ray through radio, indicates that it 
is likely to be the optical, x-ray and radio
afterglow of the gamma ray burst GRB170105.  
The  distance inferred by LIGO  to GW170104  of 
 $D_{\rm L} = 880^{+460}_{-360}$\,Mpc 
($z = 0.18^{+0.08}_{-0.07}$) leaves us with two possibilities. 

One is that  GRB170105A and its afterglow ATLAS17aeu are 
simply part of the known GRB population at higher redshift
than GW170104 and a chance coincidence in time and 2D sky area. 
This hypothesis would not violate any constraints 
from the observational data, but is not uniquely proven yet. 
The other is that ATLAS17aeu and its associated GRB is 
an unusual lower redshift transient and is physically associated
with GW17014.  Specifically, we find the following:

\begin{itemize}
\item We have detected a host galaxy, or host galaxy system 
but do not have a secure redshift. 
For the radio flux to be consistent with a relativistic 
outflow and a standard GRB, the source should be $z\gtrsim1$
(to produce a Lorentz factor $\Gamma\geq5$).  
The lack of 
Lyman-$\alpha$ forest absorption in the afterglow spectrum 
indicates a fairly secure upper redshift limit of 
 $z\lesssim 2.9$. 
The GRB is 
soft, with extended flux in the softest Konus-Wind band of 
18-70\,keV lasting about 20 seconds. Therefore a soft 
GRB lying at $1\lesssim z  \lesssim 2.9$ is not inconsistent
with any observational data, including the faint host. 
\item  We computed the probability 
that ATLAS17aeu was indeed such a chance coincidence with GW170104,  assuming it was a GRB-like event and find this to be 
 small  
 (formally $2.1-3.3\sigma$). One could argue that 
the probability calculation is somewhat selective in 
its choices of what to choose as sky rates and the 
most robust probability to quote for coincidence is 
$p=0.04$, or a chance coincidence being significant 
at $2.1\sigma$. 
\item While this is low, it is not significant enough to link 
the two events. If a redshift and distance to the host galaxy
could be established and if it were to be consistent with the 
LIGO range for GW170104, then the GRB and afterglow 
properties would not be those of the normal GRB population. 
The emitted energies in the optical, x-ray and radio 
would be similar to those predicted 
by \cite{2017ApJ...839L...7D} for plausible EM signatures from 
coalescing black holes with a 
relic circumbinary disk. It is not clear if the gamma-rays or significant time-delay as observed can be produced in this model.  
\item Hence it is essential that a distance estimate
(either spectroscopic or photometric) is measured for the 
faint host galaxy (at $r=25.6$) and its nearby companion. 
A redshift lying outside the LIGO range for GW170104 would securely rule this out as a viable counterpart. 
\end{itemize}

In conclusion, given the multi-wavelength data combined with 
the rate and probability calculations it appears
that GRB170105A and ATLAS17aeu are linked but their
energetics are most comfortably explained by a high
redshift GRB and afterglow which is just a chance coincidence 
with the position and time of GW170104.   The probability 
of such a coincidence is low, but not uncomfortably so. 

However hypothesising a
link between 
this energetic electromagnetic event and GW170104 
is quite testable.
If a soft gamma-ray to radio transient of something resembling
ATLAS17aeu is generally produced
by merging black holes (in the de Mink - King scenario), 
then we have had three events for which such optical 
coverage has been attempted (GW150914, GW151226 and 
GW170104). For the first one, issues with the skymap 
release  and observability of much of the high probability region
hindered observing. We covered finally only 4\% of the region containining the source 
\citep{2016MNRAS.462.4094S}, iPTF covered about 2.5\% \citep{2016ApJ...824L..24K}, 
and the Dark Energy Camera project covered 3\% \citep{2016ApJ...823L..33S}
hence we can discount this as a quantitative search. 
For the latter two we covered 31\% and 
43\% of the respective regions with Pan-STARRS and ATLAS 
\citep[see][]{2016ApJ...827L..40S}. 
The Dark Energy Camera and J-GEM teams covered 3\% and  29\% of 
the GW151226 map \citep[][respectively]{2016ApJ...826L..29C,2017PASJ...69....9Y}
and there was substantial overlap between our Pan-STARRS/ATLAS coverage and that of 
J-GEM.  Assuming that 
for any well observed event we have a 40\% coverage of
the skymap, then using the binomial distribution (probability of 
$d$ detections in $n$ trial events) the 
probability of getting 2 or more detections if we 
observe 8 LIGO events is 93\% (approx $2\sigma$).
The corollary is that  
to rule out optical 
electromagnetic counterparts accompanying binary 
black hole mergers at the  3$\sigma$ confidence level 
requires at least 11 events with 40\% sky probability coverage. 
This behooves the EM community to either increase the sky survey coverage (for example getting to 75\% probability coverage 
would then require only 4 events) or for LIGO and Virgo, and potentially KAGRA, to 
shrink the size of the sky maps by factors of around $\sim2$. 
It is also essential that accurate maps are released promptly 
(within 1hr of event detection) given the timescales for decline for ATLAS17aeu and the practical limitations of optical observations (e.g. daylight hours and weather).

\acknowledgments
With thank Dimitry Svinkin for insights into the publicly
available gamma ray data. 
PS1 and ATLAS are supported by the following NASA Grants
NNX08AR22G, NNX12AR65G, NNX14AM74G and
NNX12AR55G. The Pan-STARRS-LIGO effort is in collaboration with the LIGO Consortium and supported by the University of Hawaii and Queen's University Belfast.  The Pan-STARRS1 Sky Surveys have been made possible through contributions by the Institute for Astronomy, the University of Hawaii, the Pan-STARRS Project Office, the Max Planck Society and its participating institutes, the Max Planck Institute for Astronomy, Heidelberg and the Max Planck Institute for Extraterrestrial Physics, Garching, The Johns Hopkins University, Durham University, the University of Edinburgh, the Queen's University Belfast, the Harvard-Smithsonian Center for Astrophysics, the Las Cumbres Observatory Global Telescope Network Incorporated, the National Central University of Taiwan, the Space Telescope Science Institute, and the National Aeronautics and Space Administration under Grant No. NNX08AR22G issued through the Planetary Science Division of the NASA Science Mission Directorate, the National Science Foundation Grant No. AST-1238877, the University of Maryland, Eotvos Lorand University (ELTE), and the Los Alamos National Laboratory.  SJS acknowledges funding from the European Research Council under the European Union's Seventh Framework Programme (FP7/2007-2013)/ERC Grant agreement n$^{\rm o}$ [291222] and  STFC grant 
ST/P000312/1. TWC acknowledges the support through the Sofia Kovalevskaja Award to P. Schady from the Alexander von Humboldt Foundation of Germany, and thanks to T. Kr\"uhler, Y.-C. Pan and J. Graham for helping GMOS data reduction. MWC is supported by National Science Foundation Graduate Research Fellowship Program, under NSF grant number DGE 1144152. KM acknowledges support from the STFC through an Ernest Rutherford Fellowship.  Based partly on observations obtained at the Gemini Observatory under 
programmes GN-2017A-Q-23 and GN-2016B-Q-2. Gemini is 
operated by the Association of Universities for Research in Astronomy, Inc., under a cooperative agreement with the NSF on behalf of the Gemini partnership: the National Science Foundation (United States), the National Research Council (Canada), CONICYT (Chile), Ministerio de Ciencia, Tecnolog\'{i}a e Innovaci\'{o}n Productiva (Argentina), and Minist\'{e}rio da Ci\^{e}ncia, Tecnologia e Inova\c{c}\~{a}o (Brazil).
Some of the data presented herein were obtained at the W.M. Keck Observatory, which is operated as a scientific partnership among the California Institute of Technology, the University of California and the National Aeronautics and Space Administration. The Observatory was made possible by the generous financial support of the W.M. Keck Foundation.

\facilities{ATLAS, PS1, Gemini-GMOS, Keck-DEIMOS} 

\software{DOPHOT \citep{1993PASP..105.1342S}, TPHOT, IRAF \citep{1993ASPC...52..173T}, WebPIMMS/HEASARC}

\bibliographystyle{yahapj}
\bibliography{lib}

\end{document}